\documentclass[fleqn,usenatbib]{mnras}

\usepackage{newtxtext,newtxmath}

\usepackage[T1]{fontenc}
\usepackage{ae,aecompl}

\usepackage{graphicx}	
\usepackage{amsmath}	
\usepackage{amssymb}	

\usepackage{acronym}

\acrodef{AO}[AO]{adaptive optics}
\acrodef{EE}[EE]{ensquared energy}
\acrodef{PD}[PD]{photonic dicer}
\acrodef{MFD}[MFD]{mode field diameter}
\acrodef{MM}[MM]{multi-mode}
\acrodef{SM}[SM]{single-mode}
\acrodef{NIR}[NIR]{near infrared}
\acrodef{PSF}[PSF]{point spread function}
\acrodef{ULI}[ULI]{ultrafast laser inscription}
\acrodef{ELT}[ELTs]{Extremely Large Telescopes}
\acrodef{PL}[PL]{photonic lantern}

\title[Dicer]{Simulation and Optimization of an Astrophotonic Reformatter}

\author[Th. Anagnos]{Th. Anagnos$^{1}$\thanks{E-mail: tanagnos@lsw.uni-heidelberg.de}, 
R. J. Harris$^{1}$,
M. K. Corrigan$^{2}$, 
A. P. Reeves$^{3}$, 
M. J. Townson$^{2}$
\newauthor{D. G. MacLachlan$^{4}$, R. R. Thomson$^{4}$, T. J. Morris$^{2}$, C. Schwab$^{5,6}$ \& A. Quirrenbach$^{1}$}
\\
$^{1}$ Landessternwarte, Zentrum f\"ur Astronomie der Universit\"at Heidelberg, K\"onigstuhl 12, 69117
Heidelberg, Germany  \\
$^{2}$Centre for Advanced Instrumentation, Durham University, South Road, Durham DH1 3LE, UK\\
$^{3}$Deutsches Zentrum f\"ur Luft-und Raumfahrt (DLR), Oberpfaffenhofen, 
82234 We\ss ling, Germany\\
$^{4}$SUPA (Scottish Universities Physics Alliance),
Institute of Photonics and Quantum Sciences, \\Heriot-Watt University, Edinburgh, EH14 4AS, UK\\
$^{5}$Department of Physics and Astronomy,
Macquarie University, NSW 2109, Australia \\
$^{6}$The Australian Astronomical Observatory (AAO), Level 1, 105
Delhi Rd, North Ryde, NSW, 2113, Australia
}

\date{Accepted XXX. Received YYY; in original form ZZZ}

\pubyear{2017}

\begin{document}
\label{firstpage}
\pagerange{\pageref{firstpage}--\pageref{lastpage}}
\maketitle

\begin{abstract}

Image slicing is a powerful technique in astronomy.
It allows the instrument designer to reduce the
slit width of the spectrograph, increasing spectral
resolving power whilst retaining throughput. 
Conventionally this is done using bulk optics, such
as mirrors and prisms, however more recently 
astrophotonic components known as  \acp{PL} and
photonic reformatters have also been used. \\ These
devices reformat the \ac{MM} input light from a 
telescope into \ac{SM} outputs, which can then be
re-arranged to suit the spectrograph. The \ac{PD}
is one such device, designed to reduce the dependence
of spectrograph size on telescope aperture and
eliminate modal noise. \\ We simulate the \ac{PD},
by optimising the throughput and geometrical design
using Soapy and BeamProp. The simulated device shows
a transmission between 8 and 20 \%, depending upon 
the type of \ac{AO} correction applied, matching
the experimental results well. We also investigate
our idealised model of the \ac{PD} and show that the
barycentre of the slit varies only slightly with time,
meaning that the modal noise contribution is very low
when compared to conventional fibre systems. We further
optimise our model device for both higher throughput
and reduced modal noise. This device improves throughput
by 6.4 \% and reduces the movement of the slit
output by 50\%, further improving stability. This
shows the importance of properly simulating such
devices, including atmospheric effects. \\ Our work
complements recent work in the field and is essential
for optimising future photonic reformatters.

\end{abstract}

\begin{keywords}

instrumentation: adaptive optics -- instrumentation: spectrographs
\end{keywords}



\section{Introduction}
\acresetall

To detect an Earth-like planet around a Sun-like star
or an M-dwarf using the Doppler technique requires
sub-m/s radial velocity measurements.
These measurements allow us to probe the Goldilocks
zone, detecting the small planets that may harbour
life \citep[e.g.,][]{Mayor:2003, Quirrenbach:2016}.
To achieve the required precision a highly stable 
spectrograph making carefully calibrated measurements
is required. Operating at the diffraction limit, (e.g.
using a \ac{SM} fibre to feed the spectrograph) makes
this task a lot easier as the spatial profile of the
input to the spectrograph is constant with time \citep[e.g.,]
[]{Coude:1994, Crepp:2014, Schwab:2014, Jovanovic:2016}.
This is challenging, however, as a telescope rarely
produces a diffraction limited \ac{PSF}, leading
to large coupling losses. This means most current
astronomical spectrographs operate in the seeing
limited, or \ac{MM} regime and relaxing the
alignment and telescope tolerances allowing efficient
coupling of the telescope \ac{PSF}. However operating
in the seeing limited regime increases the required
size of the spectrograph. 

The dependence of the spectrograph size on the
telescope diameter feeding it can be derived
from fundamental relationships.
In its basic configuration a dispersive spectrograph
is composed of an input entrance slit into which
light is coupled from the target. This is collimated
by an optic and a dispersive element (e.g. grating
or prism) which separates the light chromatically.
Finally an optic is used to re-image the slit to the
detection plane, which measures intensity as a function
of position, and since position corresponds to
wavelength one can measure the spectrum.
The resolving power of such a spectrograph is given by

\begin{equation}
\label{res_pwr}
\hspace{3cm} R = \frac{\lambda}{\Delta\lambda} =
\frac{m \rho \lambda W} {\chi D_{\mathrm{T}}},
\end{equation}

where $\lambda$ is the central wavelength of 
observation, $\Delta\lambda$ is the smallest
wavelength difference that can be resolved, $m$ 
is the diffraction order, $\rho$ is grating ruling
density, $W$ is the illuminated grating length,
$\chi$ is the angular slit width and $D_\mathrm{T}$
is the diameter of telescope.

This relation can also be thought of as the number
of spatial modes that form a telescope \ac{PSF},
which scales with the square of the telescope
aperture $D_\mathrm{T}$ divided by the Fried seeing
parameter $\mathrm{r_{0}}$ \citep{Harris:2013,
Spaleniak:2013, MacLachlan:2017}.

If the input of a spectrograph is not diffraction 
limited (i.e. $\chi > \lambda / D_{\rm{T}} $) the
size of a given type of grating to be used in a
spectrograph depends on the telescope's diameter.
To maintain high spectral resolving power (R > 100,000)
on large telescopes, the spectrograph must also become
proportionally larger. Manufacturing errors of such
large components and difficulties stabilising their
performance make it much harder to achieve very
high measurement precision \citep{Bland-Hawthorn:2006}.

Currently, the largest telescopes have primary mirrors
around 8-10 m in diameter and require spectrographs with
meter squared dimensions, weighing many tons in order
to efficiently couple light and achieve high resolving
power \citep[e.g.][]{Vogt:1994,Noguchi:2002,Tollestrup:2012}.
The \ac{ELT} currently under construction, will be an
order of magnitude larger and a challenge for conventional
spectrograph designs \citep{Cunningham:2009, Mueller:2014, Zerbi:2014}.

To reduce the size of the instrument, the number of 
modes can be reduced using \ac{AO}. In particular
extreme \ac{AO} systems can deliver a close to perfect
diffraction limited \ac{PSF} (> 90 \% Strehl ratio)
in the H-band, though these are limited by a narrow
field of view and require a very bright guide star
\citep[e.g.,][]
{Dekany:2013,Agapito:2014,Macintosh:2014,Jovanovic:2015}.
Not all telescopes are equipped with an extreme \ac{AO}
system that can provide a high-Strehl \ac{PSF}, and they
cannot provide this level of performance at visible
wavelengths.

For non diffraction-limited systems another
approach to reduce size is spatial reformatting
of the coupled target into a slit geometry, commonly
known as image slicing \citep[e.g.,][and references
therein]{Weitzel:1996}. The input can be manipulated
and smaller segments can then be fed to smaller, more
stable instruments \citep{Allington-Smith:2004,
Hook:2004, Harris:2013}.

Astrophotonic examples of this technique include PIMMS
(the Photonic Integrated Multimode Micro Spectrograph)
\citep{Bland-Hawthorn:2010}, an \ac{ULI} device in
conjunction with a multicore fibre \citep{Thomson:2011},
the Photonic TIGER concept which is a multicore fibre
feeding a spectrograph \citep{Leon-Saval:2012}, and the
\ac{PD} a \ac{ULI} photonic spatial reformatter
\citep{Harris:2015}. They are all composed of a combination
of optical fibre guided-wave manipulations and transitions,
which were developed from the \ac{PL} \citep{Leon-Saval:2005,
Leon-Saval:2013, Birks:2015}. The device converts the
\ac{MM} \ac{PSF} from the telescope to many \ac{SM}
inputs to feed the spectrograph \citep{Cvetojevic:2009,
Cvetojevic:2012}. Initially \acp{PL} were developed 
using fibres \citep[e.g.,][]{Yerolatsitis:2017},
but later other groups manufactured them as integrated
devices using different techniques \citep[eg.][]{Thomson:2011,
Spaleniak:2013}.

Potentially one of the largest advantages of working
in the \ac{SM} regime is the elimination of modal
noise in the spectrograph, allowing more precise
calibration \citep{Probst:2015}. Modal noise is caused
by the temporally varying \ac{MM} input to the 
spectrograph, resulting in a change of the measured
barycentre for a given wavelength. This translates 
into spectrograph noise and thus is a major limiting
factor for precise spectroscopic measurements using
\ac{MM} fibres \citep[e.g.,][]{Lemke:2011,Perruchot:2011,
McCoy:2012,Bouchy:2013,Iuzzolino:2014,Halverson:2015}.
A single mode fibre acts as spatial
filter eliminating the modal noise as only the
fundamental mode propagates (neglecting polarisation)
and higher order modes radiate out in the cladding.
Using reformatters has been proposed to combine
the throughput of a \ac{MM} system with the modal noise
free behaviour of a \ac{SM} fibre, though recent papers
have shown that the optical configuration should be treated
carefully for parts bringing in modal noise causing
the final system to not be modal noise free 
\citep{Spaleniak:2016,Cvetojevic:2017}. Finally, it
should be mentioned that astrophotonic reformatters
do not preserve imaging information as a conventional
image slicer does.

\par In this paper, we compare the simulated performance
of the \ac{PD}, a photonic reformatter tested on-sky by
\cite{Harris:2015} with computer models. This astrophotonic
spatial reformatter re-arranges the coupled \ac{PSF} 
into a diffraction-limited pseudoslit output. It has
the potential to enable more precise high-resolution
spectroscopic measurements of astronomical sources,
if it can be shown to be a modal noise free design.

In Section \ref{sec:methods} we describe the 
configuration parameters taken into account for
the simulated version of the \ac{PD}. Then we present
results in Section \ref{sec:results}, including 
the procedure followed and the techniques used for
the optimisation. We discuss the results
in Section \ref{sec:discussion} and conclude in
Section \ref{sec:conclusion}.

\section{Methods} 
\label{sec:methods}

In order to calibrate future designs, and test their
potential, realistic simulation conditions are required.
For this work two tools were combined to simulate the
\ac{PD}'s on-sky performance: Soapy \citep{Reeves:2016},
a Monte Carlo \ac{AO} simulation program, is used to model
the atmosphere and its impact on the performance of the
\ac{PD}; and the finite-difference beam propagation 
solver BeamPROP by RSoft \citeauthor{rsoft}, which is
used to model the \ac{PD} itself. 

The simulations were performed in two ways: First, 
Soapy was used to determine an \ac{AO}-corrected
output phase, which could then be used as an input
for the BeamProp software, and secondly using the
on-sky data from \cite{Harris:2015} as the
input (real). In order to identify areas of
improvement, these two methods are compared.

\subsection{Soapy Configuration}
Soapy was configured to approximate the CANARY 
\citep{Myers:2008} parameters used on-sky for the
\ac{PD} tests (see Table \ref{soapy_set}). The
simulation was run in the same three \ac{AO} modes
as used on-sky, namely closed-loop, tip-tilt
and open-loop. To match the on-sky
\ac{AO} performance the seeing parameter (Fried
parameter - $\mathrm{r_{0}}$) was set to a range
of 0.09 to 0.11 m, which is representative of
the conditions encountered during the on-sky
experiments described in \cite{Harris:2015}.
In the first step, Soapy is used
to produce 12000 \ac{NIR} data frames, each with
an exposure time of 6 ms. The science camera 
parameters of the Soapy output frames were 128x128 
pixels, covering 3.0 arcseconds, just under ten 
times the angular size of the \ac{PD} on-sky. 
Unlike the on-sky camera data, these frames contain
both phase and amplitude information, which was 
found to be essential to the simulation accuracy
and is detailed in Section \ref{sec:throughput}.
These Soapy frames were used as an input to BeamProp.

\begin{table*}
 \caption{Simulation Soapy input parameters}
 \label{soapy_set}
 \centering
 \begin{tabular}{l c c c}
 \hline\hline
  & & Modes of \ac{AO} operation\\
 \hline
 & closed-loop & open-loop & tip-tilt \\
 \hline \hline
             Parameters \\
 \hline
   Seeing (arcsec) & 1.03 & 0.94 & 1.15 \\
   Instantaneous Strehl ratio (mean) & 0.26 & 0.08 & 0.07  \\
   Long exposure Strehl ratio (mean) & 0.12 & 0.01 & 0.01  \\
   Fried parameter $\mathrm{r_{0}}$ (m) (@1550 nm) & 0.1 & 0.11 & 0.09 \\
   Atmosphere layers & 5 & 5 & 5 \\
   DM integrator loop gain tip-tilt & 0.3 & 0.001 & 0.3\\
   DM integrator loop gain Piezo & 0.3 & 0.001 & 0.001\\
   \hline\hline
   \end{tabular}
    \label{table:soapy}
\end{table*}

\subsection{BeamProp Configuration}

\begin{figure}
\centering
 \includegraphics[width=0.5\textwidth]{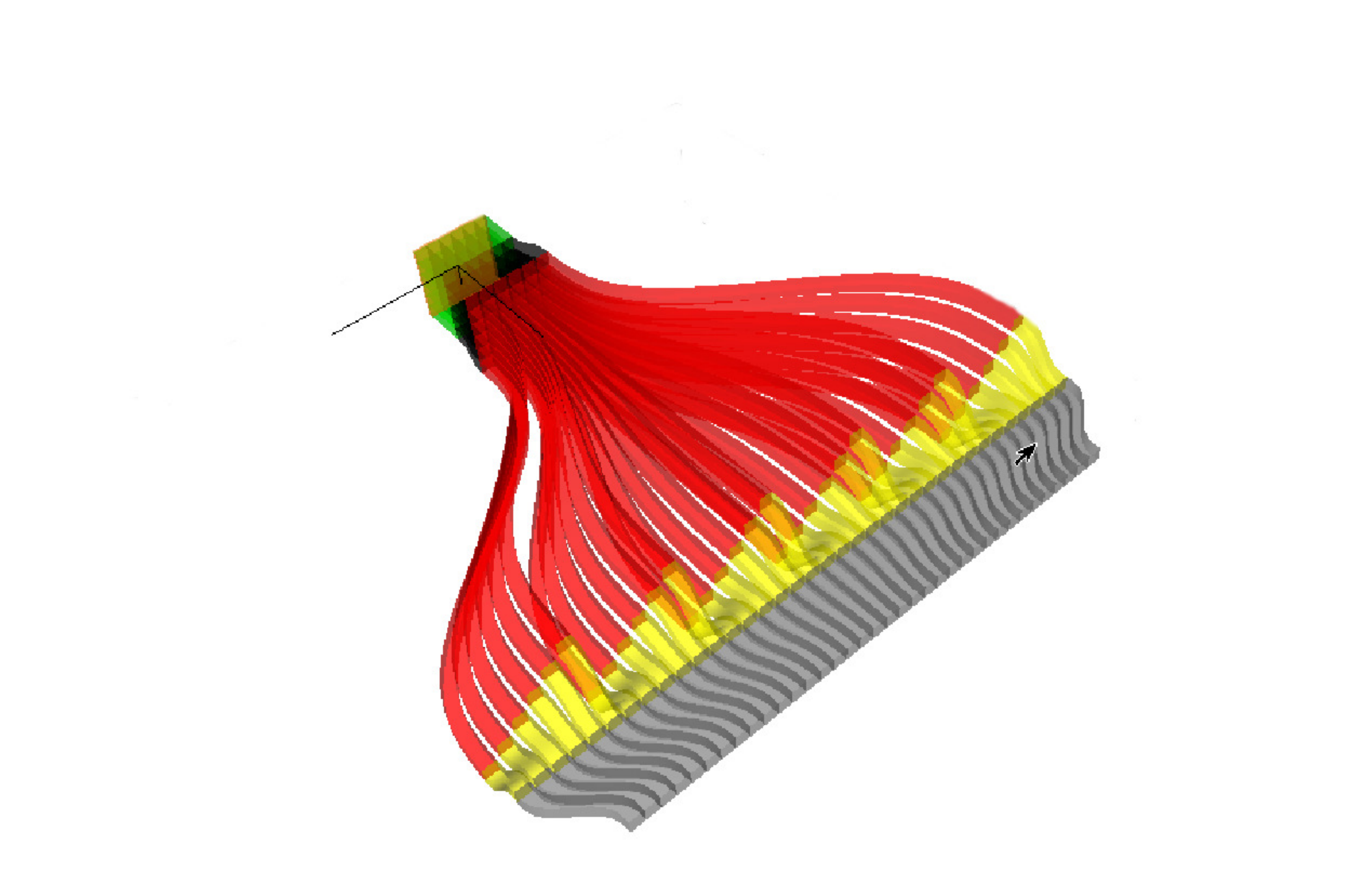}
 \caption{The Photonic Dicer 3D design in the RSoft CAD
 environment. The colours indicate the 5 different
 transition planes used.}
 \label{fig:rsoft}
\end{figure}

Each frame from Soapy was then used as an input for 
BeamProp; the angular size of the \ac{PD} was set to
321 mas. For these simulations the \ac{PD} architecture
was as described in \cite{MacLachlan:2014} and is shown
in Figure \ref{fig:rsoft}. BeamProp requires refractive
indices for both the core and cladding of the device.
The cladding is a borosilicate glass (Corning, EAGLE200),
which has a refractive index $\mathrm{n_{cl}}$ of 
$\sim$1.49 at 1550 nm. As no  refractive index measurements
were made of the waveguides in the \ac{PD}, this value
is taken from \cite{Thomson:2011}. The value
$\mathit{\Delta} = \frac{n_{\mathrm{core}}-n_{\mathrm{clad}}}
{n_{\mathrm{core}}}\approx 1.76\times 10^{-3}$, is expected
to be close to the waveguides in the \ac{PD}, but due to
differences in the inscription parameters, small variations
are expected (see Table \ref{table:ULI}).

By default, BeamProp does not take into account the
material propagation loss. For our simulations, we
ran tests using losses of 0.1 dB/cm \citep{Nasu:2005},
though this was shown to be small in comparison to the
losses due to geometrical changes (< 2 \% over the \ac{PD}
length). However, this will need to be taken into account
in future modelling with more efficient devices.

\begin{table}
 \centering
 \caption{Comparison of ULI inscription parameters
 used in \citeauthor{Thomson:2011} and 
 \citeauthor{Harris:2015}}
 \begin{tabular}{lcc}
 \hline
 Parameters & Thomson et al. & Harris et al.\\
 \hline
 $n_{\mathrm{cl}}$ (@1550 nm) & $\sim$ 1.49& $\sim$ 1.49\\
 Pulse Energy (nJ) & 165 & 251\\
 Pulse repetition rate (kHz) & 500 & 500\\
 Pulse duration (fs) & 350 (1047 nm) & 460 (1064 nm)\\
 \hline
 \end{tabular}
 \label{table:ULI}
\end{table}

\begin{figure}
 \begin{tabular}{ll@{}}
 \includegraphics[width=0.24\textwidth]{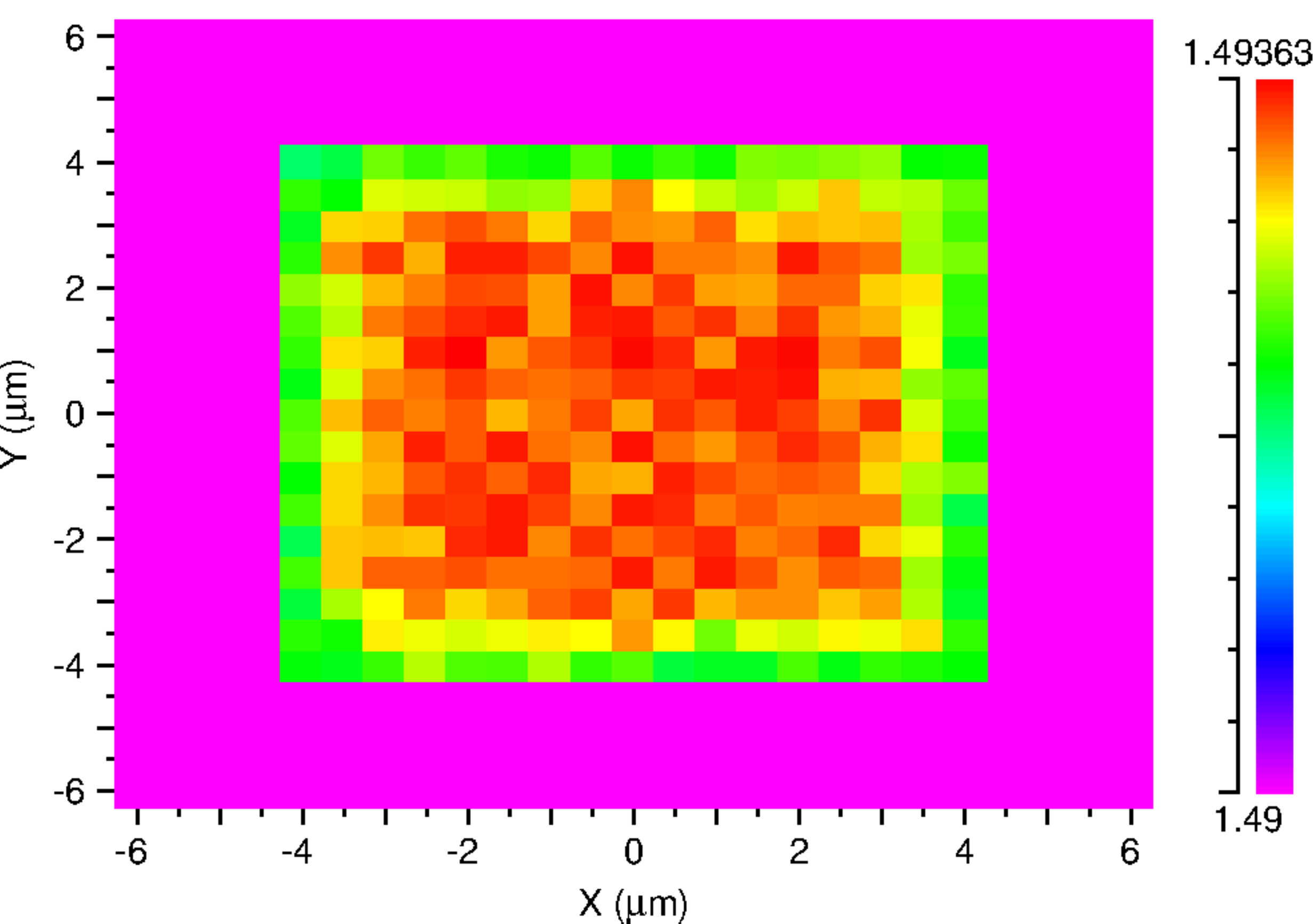} 
 & \includegraphics[width=0.24\textwidth]{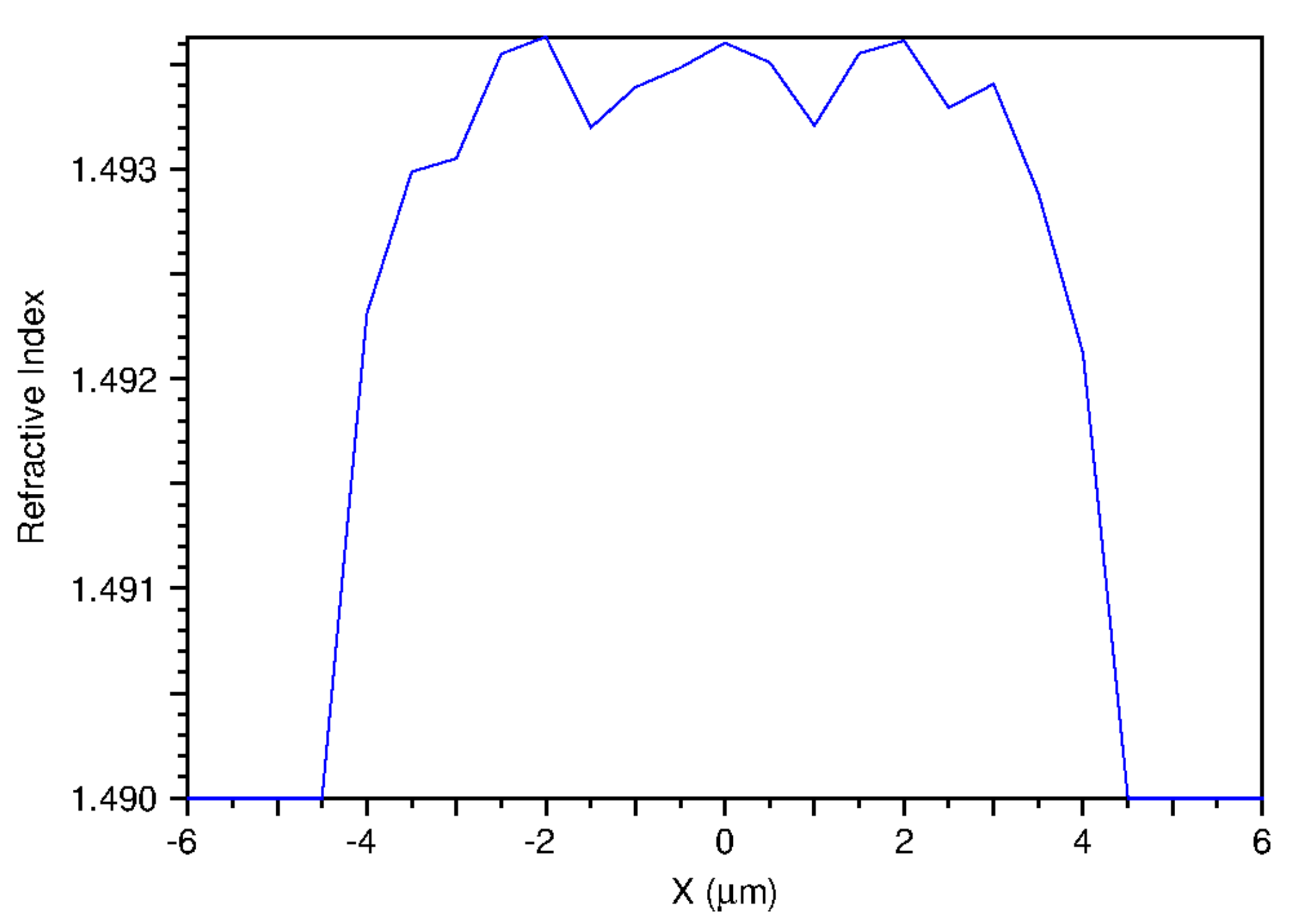}
 \end{tabular}
\caption{\textbf{Left}: Colour map showing the
refractive index profile of a noisy waveguide,
\textbf{Right}: Cross-section of the colour map.}
\label{ref_prof}
\end{figure}

To increase the accuracy of the simulations, introducing
noise to the step refractive index profile of the
waveguides was considered, similar to that measured
by \cite{Thomson:2011} (see Figure \ref{ref_prof}). 
This greatly increased simulation time and the 
differences in efficiency 
between noisy and noiseless waveguides were found to
be minor (< 0.001 \%). Thus simulations were performed
without taking into account noise in the refractive index
profile of the waveguides.

\subsection{Throughput calculation}
In order to calculate the total throughput
($T_{\mathrm{tot}}$) of the \ac{PD} the
ratio of the flux in the slit output ($F_{\mathrm{slit}}$)
to that of the input field ($F_{\mathrm{ref\star}}$)
was taken for each of the science frames. As BeamProp
does not take into account any size differences in
images, a constant $k$ is used to normalise the input
and output spatial sizes of the fields as they were
different; this results in

\begin{equation}
\centering
\label{eq3}
\hspace{2cm} {T_{\mathrm{tot}} = \frac{F_{\mathrm{slit}}
(i)}{F_{\mathrm{ref\star}}(i) \times k} ,
\quad i = \# frames}.
\end{equation}

\subsection{Dicer Plane Optimisation}

The \ac{PD} was designed in 2013, without the ability
to do the full system modelling available using our 
software suite. This means that there are potential 
optimisation possibilities that were not taken into
account. To investigate this, we use a Monte Carlo
simulation routine built into BeamProp to calculate
the relative losses for different propagation planes (see Figure 
\ref{fig:rsoft}), 
changing the size of the \ac{PD} to the optimal one.

In order for the transitions to have low losses, they 
should be gradual \citep{Birks:2015}. However, as 
using \ac{ULI} results in relatively high material
and bend losses these transition losses need to be
balanced against length. 

Simulation results for the optimal device (see Figure
\ref{partial_ee}) show that the optimal
\ac{PD} length is shorter than the constructed
one by several mm, leading to greater throughput and a more
compact design.

\section{Results}
\label{sec:results}

\subsection{Throughput performance results} \label{sec:throughput}
Here, the throughput results are presented from 
the simulation configurations as described in 
section \ref{sec:methods}. As stated above, the
Soapy \ac{AO} modes were configured to approximate
the on-sky corresponding performance. Consequently,
the tip-tilt \ac{AO} mode was adjusted to perform
worse than the open-loop case, in terms of correction,
by regulating the seeing/Fried parameter in our
simulations (see Table \ref{table:soapy}).
Hence, simulations were performed using our produced 
Soapy data (phase and amplitude information provided)
and real on-sky images acquired in the focal plane
at the input of the \ac{PD} provided by CANARY 
\citep{Myers:2008}. As Canary uses an InGaAs camera
only intensity is recorded, therefore for the 
simulations a flat phase front (all phase = 0) and
the square root of the intensity (amplitude) is used.

\begin{figure}
\centering
\includegraphics[width=0.5\textwidth]{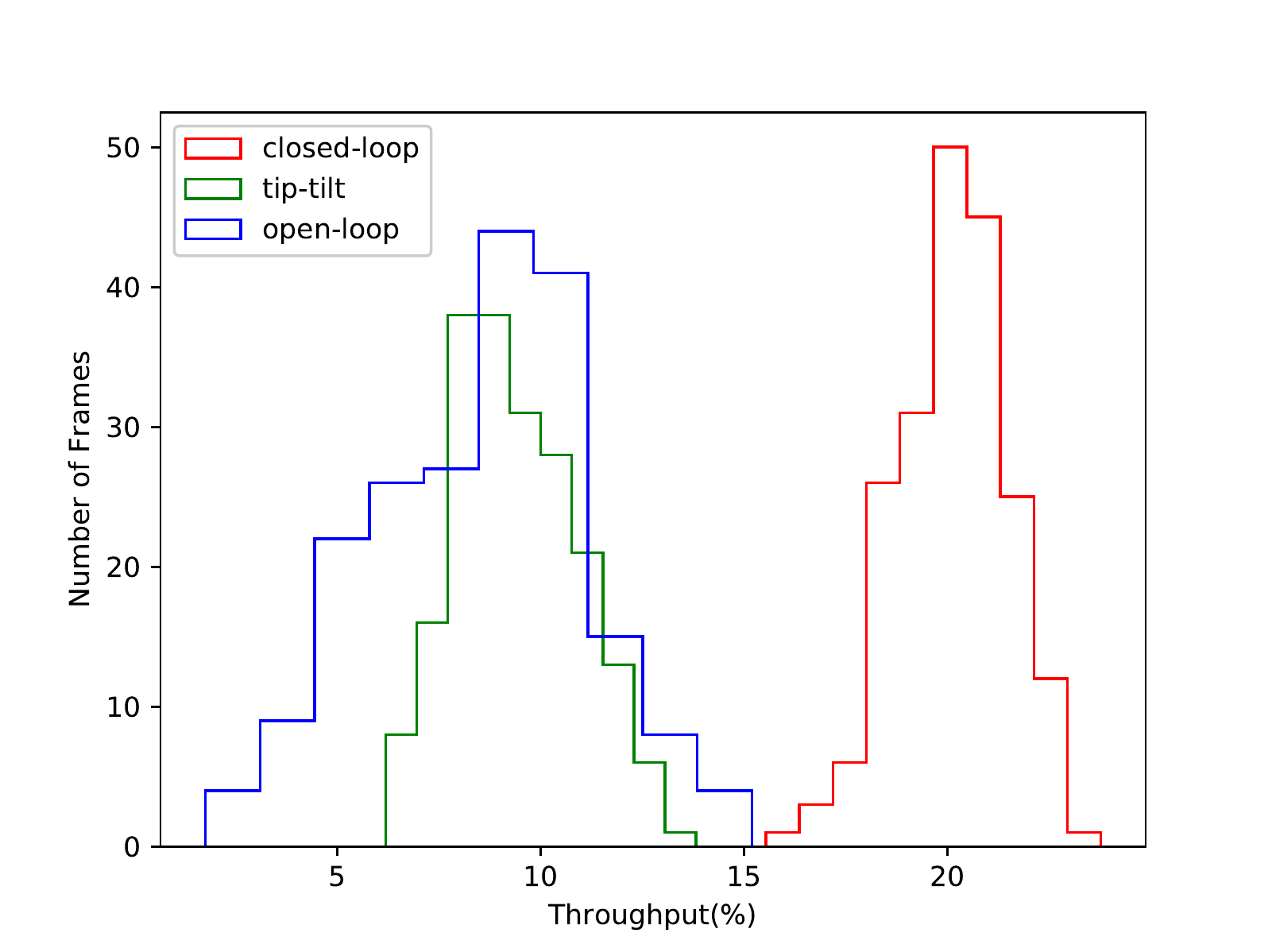}
\caption{Histogram plot of throughput measurements
in the three different AO modes, with each mode
containing 12000 simulation results binned by a
factor of 60. In order of correction, red shows
closed-loop, green tip-tilt correction and blue
shows open-loop.}
\label{hist_inst}
\end{figure}

The results of simulating 12000 frames are shown 
in Figure \ref{hist_inst}. For closed-loop operation
mode (full \ac{AO} correction) the transmission of the
\ac{PD} was measured to be 20 $\pm$2 (\%). In
open-loop operation mode the transmission was
measured to be 8 $\pm$2 (\%); and for tip-tilt 
correction results shown to be 9 $\pm$2 (\%).

The camera data taken from the on-sky run (real)
were also simulated by BeamProp and the results
are shown in Table \ref{table:res}. This shows an 
overestimation of the throughput by a factor of
$\sim$2. The reason of this overestimated result
is the absence of phase information in the on-sky
data fields and as a consequence BeamProp considers
zero-phase everywhere.

\begin{figure}
\centering
\includegraphics[width=0.5\textwidth]{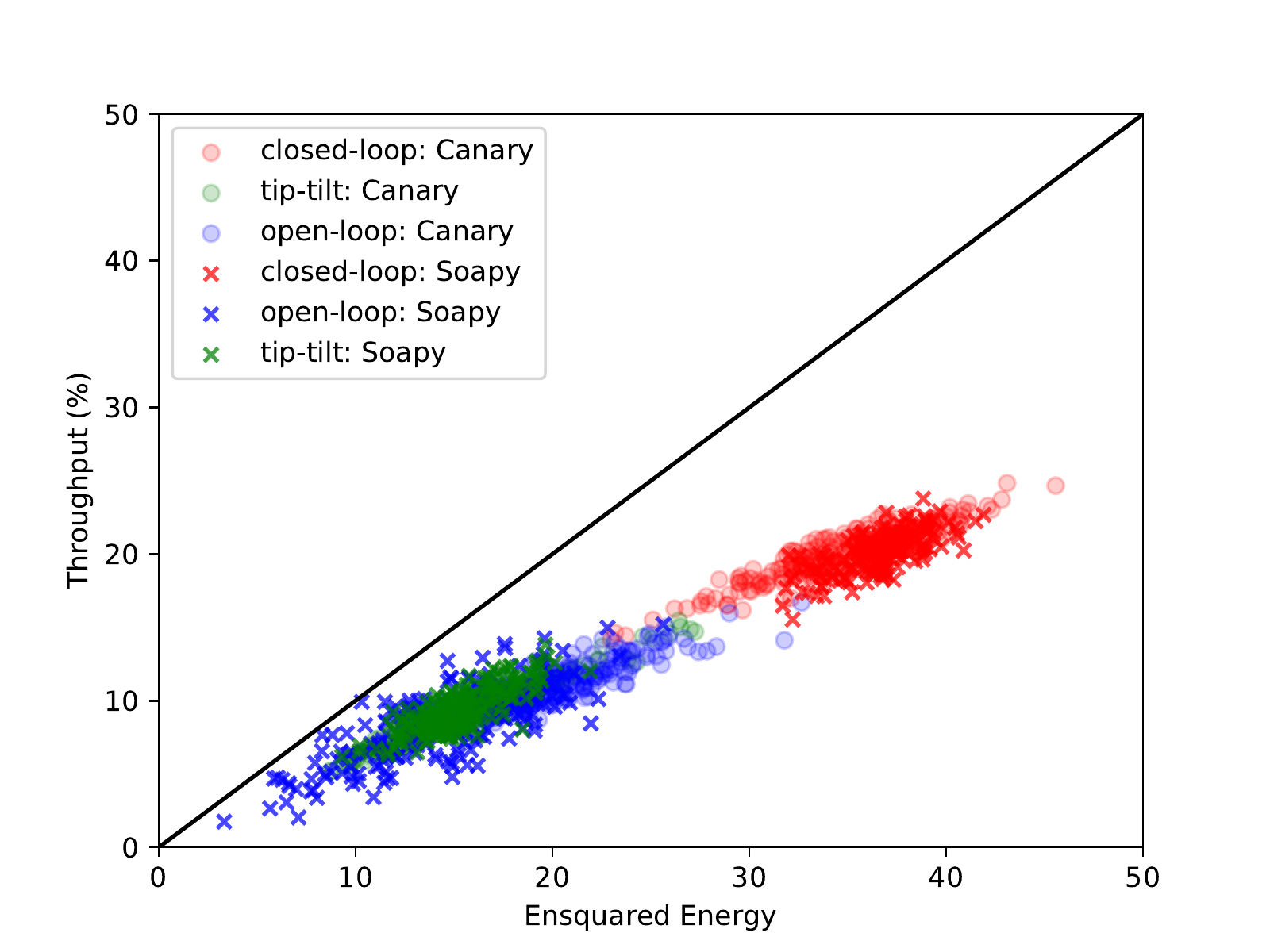}
\caption{Throughput measurements of the simulated 
\ac{PD} slit end versus the amount of light coupled 
at 321 mas (the square entrance of the device as 
configured in BeamProp). This is shown for the 
simulated data of Soapy (bold colour) and on-sky 
results (transparent) in all three \ac{AO} 
operating modes. Note that the number of points
are binned by a factor of 60 into 200 points
from the 12000, for each \ac{AO} mode.}
\label{th_inst}
\end{figure}

As with \cite{Harris:2015} we also investigated the
ratio of output power in the slit to input power
coupled to the \ac{PD}, in order to calculate a value
of light transmitted through the \ac{PD}. To do this
the \ac{EE} at the input of the \ac{PD} was calculated
and plotted against the corresponding throughput.
Figure \ref{th_inst} shows the result of this; as in 
\cite{Harris:2015} we see a positive linear correlation
of \ac{EE} with calculated slit output power. The
black line shows where the input \ac{EE} and output
throughput are equal. Some values are close to equal;
this is due to evanescent field coupling which is
further explained in section \ref{sec:ev_coupling}.

\begin{table}
	\centering
	\caption{Fractional throughput results comparing
    theoretical simulations and on-sky conditions. The
    incorrect results for real measured input data + 
    BeamProp show a factor of two overestimation because
    BeamProp assumes zero phase if phase information is
    not provided; this highlights the importance of having
    phase information of the input beam in the simulations.}
	\label{table:res}
	\begin{tabular}{lccc}
		\hline
        &&Data and results&\\
        \hline \hline
		\ac{AO} mode & On-sky & 
        Soapy & On-sky\\
         & & +BeamProp & +BeamProp\\
		\hline
        closed-loop (\%) & 20 $\pm$ 2 & 20 $\pm$ 2 & 45 $\pm$ 2\\
		tip-tilt (\%) & 9 $\pm$ 2 & 9 $\pm$ 2 & 20 $\pm$ 2\\
        open-loop (\%) & 11 $\pm$ 2 & 8 $\pm$ 2 & 24 $\pm$ 2\\
        \hline
	\end{tabular}
\end{table}

\begin{figure}
	\centering
		\includegraphics[width=0.5\textwidth]{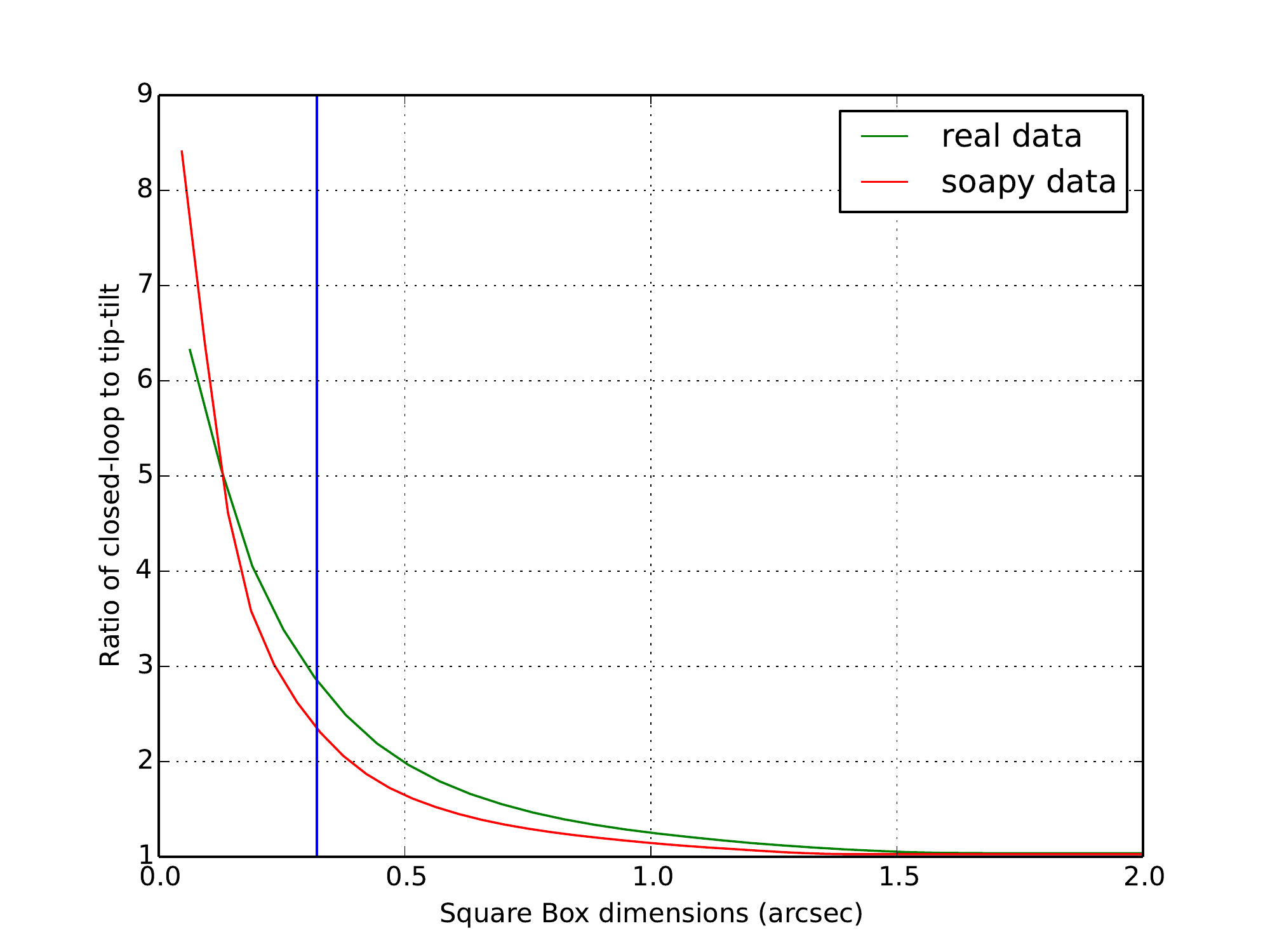}
		\caption{The relation of closed-loop and tip-tilt
		\ac{AO} mode ratios of \ac{EE} as a function of spatial
        scale (square box centred), plotted for both simulated
        (Soapy) and on-sky averaged data (real). The \ac{PD}
        square entrance size is represented by the blue
        vertical line.}
	\label{ratio_ct}
\end{figure}

For a better understanding of the coupling efficiency
\ac{EE}, the ratio between closed-loop and tip-tilt
correction was calculated and plotted versus the
device \ac{MM} entrance input size for averaged
Soapy and real data images (Figure \ref{ratio_ct}).
This figure illustrates that the \ac{EE} under 
closed-loop mode is higher than that of tip-tilt
by a factor of $\sim$ 2.8 for real and
$\sim$ 2.4 for Soapy data. This factor varies
inversely with the spatial size of the sampling
as a function of overall throughput.

\subsection{Optimisation results}

\begin{figure}
	\centering
		\includegraphics[width=0.5\textwidth]{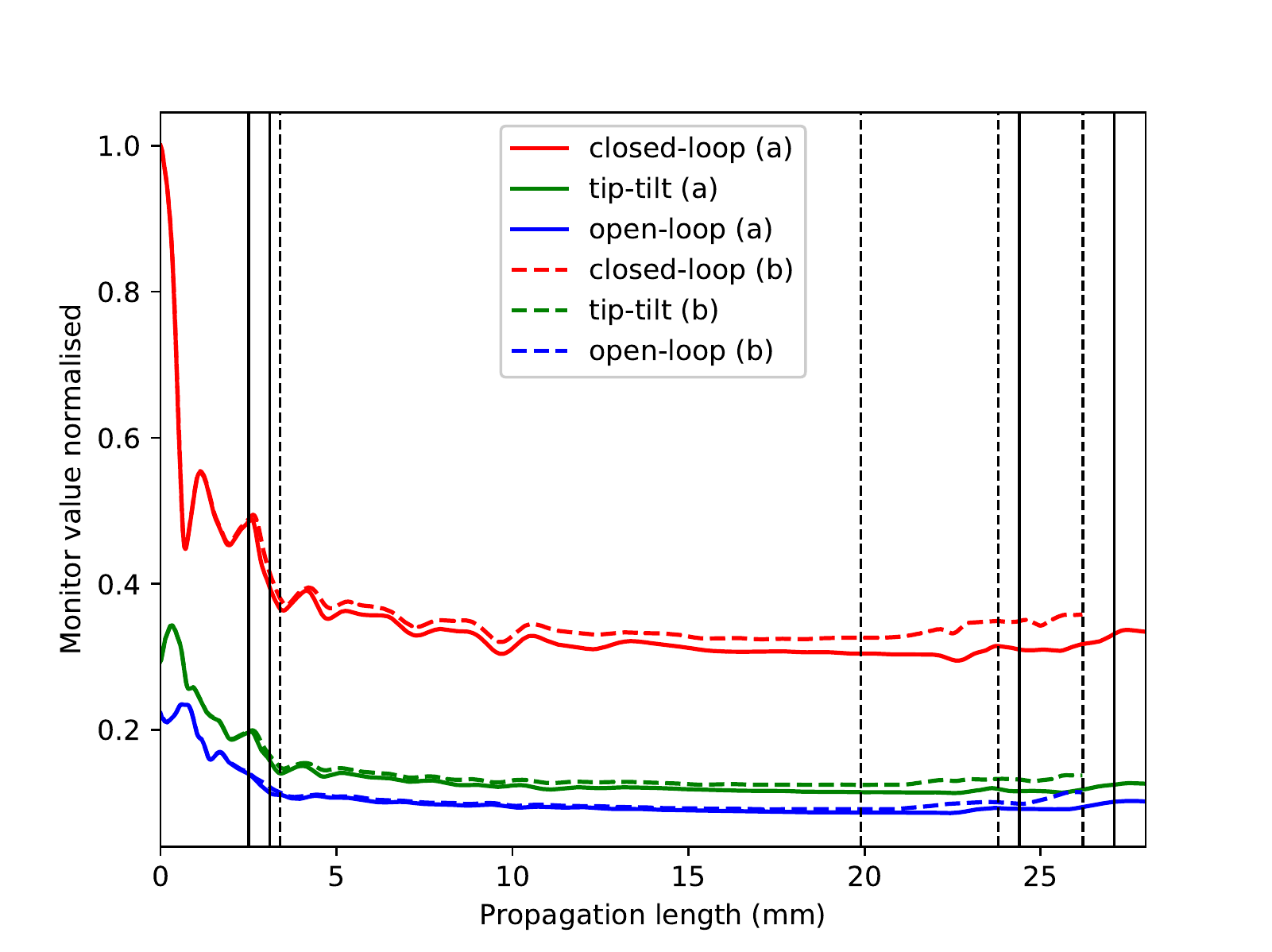}
		\caption{Co-added computed power enclosed inside
		the 36 waveguides as a function of the propagating
		length. Vertical solid black lines indicate
		the 5 transition planes of the device as it was
		originally built, and vertical dashed black lines 
		represent the optimised locations of the planes.
		This is shown for the three \ac{AO} operating modes in
		three different colours (constant lines for default
		\ac{PD} (a) and dashed lines (b) for the optimised
		version correspondingly). Computed powers are
		normalised according to the maximum of each \ac{AO} 
        modes. Averaged frames of all three \ac{AO} modes
        were used as an input. Power fluctuations are 
        discussed in section \ref{sec:ev_coupling}.}
	\label{partial_ee}
\end{figure}

In order to optimise the \ac{PD}, the average of 
the real and imaginary parts of the electric field
of the frames from a closed-loop dataset by Soapy
was chosen. Using this as an input, a Monte Carlo
simulation was performed on the \ac{PD}, optimising
each of its transition planes for throughput by
scanning for different lengths among the 5 transition
planes of the device. The results of this are shown
in Figure \ref{partial_ee}. In this figure, throughput
results from simulations with the optimised and
unoptimised versions of the \ac{PD} using all of the
three \ac{AO} modes as an input are plotted against
the propagation length of the device. The solid and
dashed lines represent the unoptimised and optimised
\ac{PD}, respectively. In this illustration we notice
the shorter more efficient version of the \ac{PD}, as 
well as the high coupling losses at the entrance
input of the device, where the \ac{PL} section is
located. That means the transition can be further 
improved to be more adiabatic and thus lower in loss.

\subsection{Modal noise results}
\label{sec:mnoise}
To investigate whether our theoretical \ac{PD} was
subject to modal noise, we performed two analyses.
The first is similar to a classical modal noise
experiment, where the measured barycentre of the
slit moves \citep{Rawson:1980,Chen:2006}. To do 
this we chose a single wavelength and examined the
stability of the near field image of the slit using 
our Soapy produced images as an input. The second
is a more recently discovered phenomenon, namely 
periodic variations of throughput as a function of
wavelength, due to modal mismatch in the reformatting
devices \citep{Spaleniak:2016,Cvetojevic:2017}.

\begin{figure}
\centering
\includegraphics[width=0.5\textwidth]{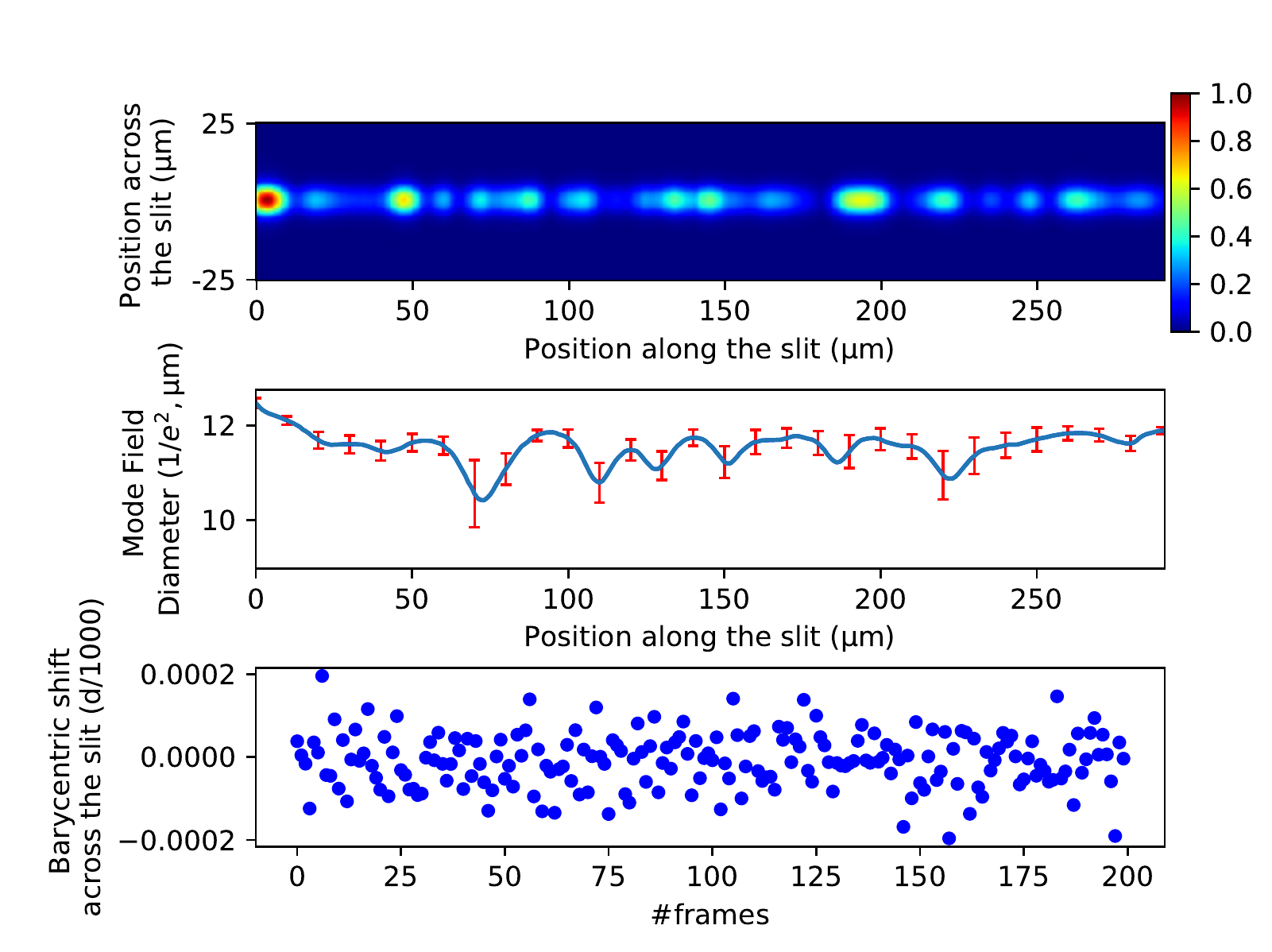}
\caption{Top panel: Near field averaged image 
(intensity) of the slit from BeamProp simulations
(@1550 nm). Middle panel: MFD of the slit profile including 
1$\upsigma$ errors from individual frames. Bottom
panel: Near field barycentre shifts across the slit.}
\label{slit_com}
\end{figure}

To check the stability of the slit, the intensities
of output frames from the simulations were averaged. 
The variation of the \ac{MFD} and its barycentric
position were calculated to look for disturbances
of the coupled field that are translated to a 
different speckle pattern at the slit output.
Figure \ref{slit_com} presents the analysis results
of the \ac{PD}. In the top panel the averaged slit image
(intensity) from BeamProp is illustrated. The middle
panel shows the \ac{MFD} of the slit profile calculated
from the Gaussian fit, and the bottom panel depicts
the barycentre position of the \ac{MFD} calculated across
the slit. Measurements of the barycentre movement 
are presented as a portion of one-thousandth of the
core diameter ($d$/1000). Results show a mean variation
of 1.2 $\upmu$m (10\% of the averaged \ac{MFD}) in the \ac{MFD}
dimension, while the semi amplitude barycentre variation
was found to be of the order of $2\times 10^{-4}$ ($d$/1000).
It should be noted that the simulations did not include
any manufacturing errors in the straightness of the
slit. These variations degrade the spectral resolving
power and introduce noise and uncertainties in the 
produced spectra.

\begin{figure}
\centering
\includegraphics[width=0.5\textwidth]{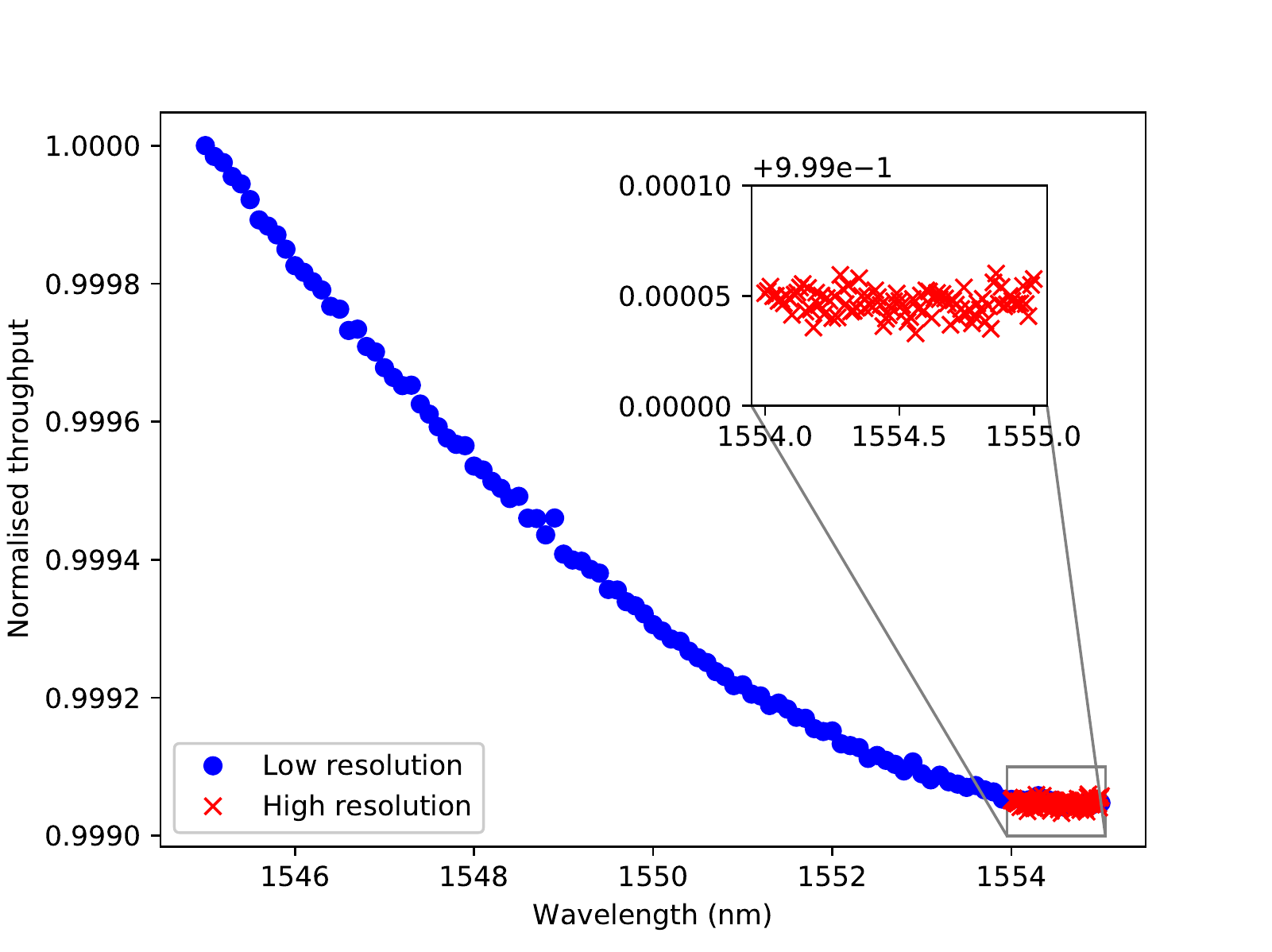}
\caption{Throughput performance of the \ac{PD} as 
a function of wavelength for low spectral
resolution scanning with steps of 0.1 nm (R $\sim$ 
15.500), and high spectral resolution scanning with 
steps of 0.01 nm (R $\sim$ 155.000). \textbf{Inset:} 
High spectral resolution magnified.}
\label{modal_res}
\end{figure}

Measurements of the throughput were performed in
two wavelength regimes; in the first one covering
the 1545 - 1555 nm wavelength range with 0.1 nm steps to
approximate a typical low resolution spectrum
(R $\sim$ 15.500), and in the second one covering the
1554.5 - 1555.5 nm wavelength range with 0.01 nm
steps corresponding to a typical high resolution
spectrum (R $\sim$155.000). It should be noted that
the launch mode profile remained the same in those 
simulations for all wavelengths, namely a 50 
$\upmu$m (MFD @ $\mathrm{1/e^{2}}$) representative
of a diffraction-limited input injected into the 
entrance of the \ac{PD}. Normalised throughput
results are presented in Figure \ref{modal_res}, where
it can be seen that there is no significant variation
of throughput with wavelength, both for high and low
resolution simulations.

\section{Discussion}
\label{sec:discussion}

\subsection{Adaptive optics performance}

In order to match the performance for each \ac{AO} 
operation mode, the datasets from Soapy were compared
to the corresponding on-sky ones. By comparing the
\ac{EE} within a growing box starting from the 
centre of averaged data frames as a function of 
square box spatial dimensions, we matched our 
simulated to on-sky ones. We found most results
converged for the same \ac{AO} parameters as on-sky,
though the mean seeing value of all \ac{AO} modes
used in Soapy was 1.04 arcseconds instead of the
0.7 arcseconds as seen on-sky (see Table
\ref{table:soapy} \& \cite{Harris:2015}). 
This might be caused by various factors, including
the unstable atmospheric conditions on-sky,
vibrations due to electronics in the telescope and
the impact of the wind on the telescope dome and 
around its components. This raises the question of
how best to optimise future simulations and what
data to take for future on-sky tests. Future work
will involve adding more noise to our simulations 
to try to better compare our results with on-sky 
data. It should be noted also that the effect of
changing atmospheric conditions was considered in
order to represent better the on-sky conditions
(see Figure \ref{partial_ee} and Table \ref{table:soapy}
by adjusting the seeing parameter in each \ac{AO}
mode).

\subsection{F-ratio calibration}

\cite{Harris:2015} state that the relative
scaling between the calibration and main arms of
their experiment configuration had a magnification
mismatch. This was caused by errors in focal length
calculation due to the extremely short focal length
$\sim$ 4.5 mm lenses that imaged the \ac{PSF} 
generated by CANARY onto the \ac{PD} entrance. Our
initial tests were performed with their platescale
of 7.96 arcseconds/mm (a \ac{PD} entrance aperture
of 405 mas), which led to an underestimation of the
on-sky throughput performance. Following further
investigation we concluded that a platescale of 6.37
arcseconds/mm (\ac{PD} entrance aperture of 321 mas)
produced a much better fit of the resulting throughput
compared to the on-sky results. With the appropriate
corrections on magnification, we found that their
results fit ours. As their lenses had short focal
lengths it is likely that their scaling has large
errors, which leads to the mismatch. In future
on-sky experiments it would be extremely useful to
have accurately characterised optical designs.

\subsection{Evanescent field coupling} \label{sec:ev_coupling}

In Figure \ref{th_inst} we see that the measurements
with lower \ac{EE} (and hence less light into the 
\ac{PD}) show a throughput closer to the input \ac{EE}
(a higher device transmission); while when the \ac{EE}
was increased, the fraction of light passing through
the \ac{PD} appears to drop. 

\begin{figure}
\centering
\includegraphics[width=0.5\textwidth]{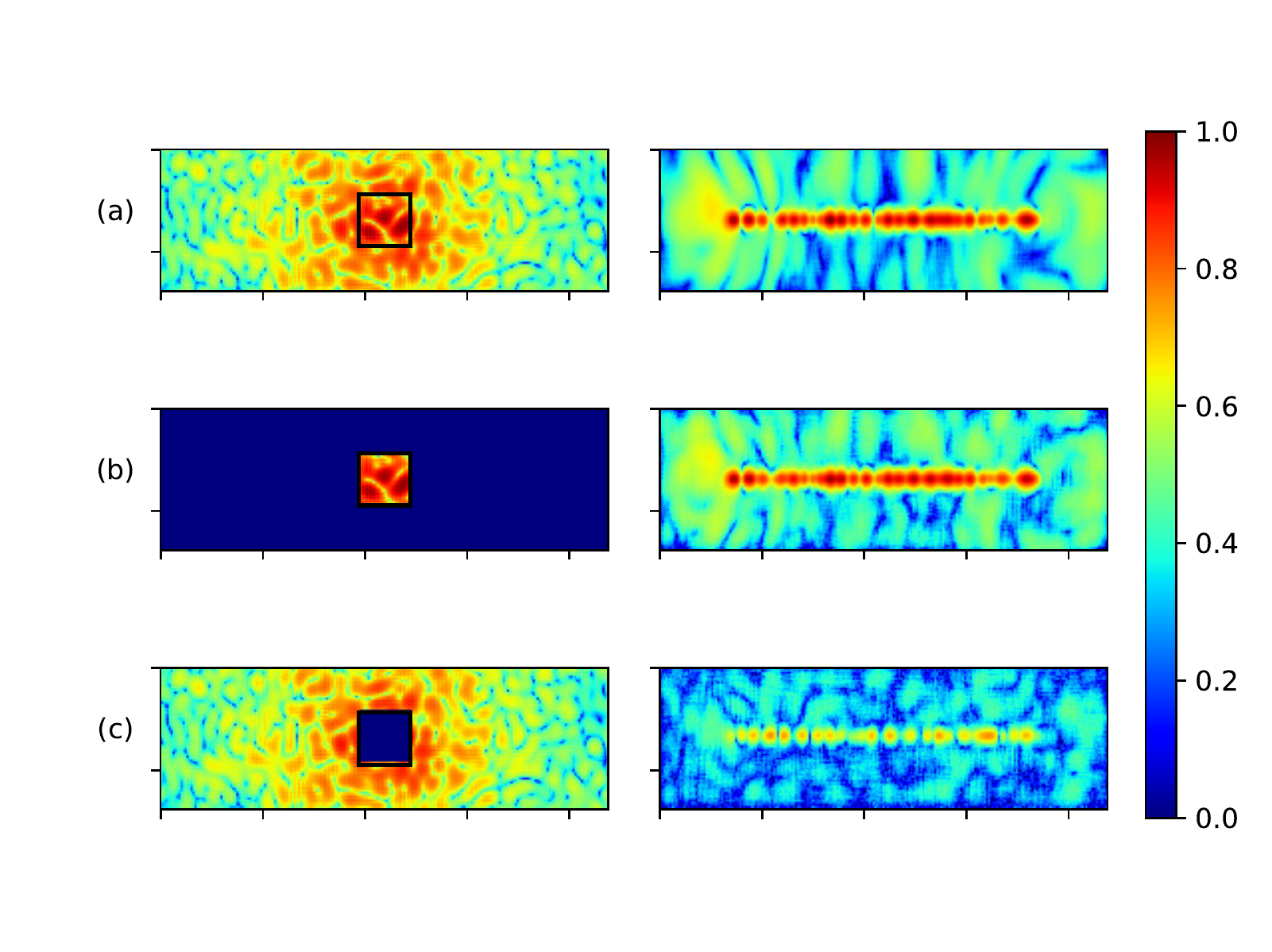}
\caption{(a-c) Colour map images (logarithm of intensities)
of the Soapy closed \ac{AO} mode input data at the left
together with the corresponding reformatted output of
the \ac{PD} at the right, for 3 different \ac{EE} 
coupled to device spatial simulation domain, (a - full
field) and (b - cut field) 42.05 \% of the full frame,
(c - cut-inside field) 57.94 \% of the full frame. 
The simulated spatial domain for each of the 6 frames is
438$\times$138 $\upmu$m.}
\label{EE_couple_cl}
\end{figure}

To investigate this effect, a test was conducted
with three data frames from Soapy, one in closed-loop
mode, one in open-loop and one in tip-tilt (full field).
As with our other simulations, this was propagated 
through the \ac{PD} and the throughput measured. The 
field outside the \ac{PD} was then set to zero and the
simulation was re-run (cut field). A third simulation
was then performed with the field inside the \ac{PD}
set to zero (Cut-inside field) (see Figure \ref{EE_couple_cl}
(full, cut, cut-inside field)). 

To calculate the relative throughput for each simulation
per \ac{AO} mode, we use the following equation

\begin{equation}
\begin{split}
T_{\rm{tot}} & = EE_{\rm{b}} \times T_{\rm{b}} + EE_{\rm{c}} \times T_{\rm{c}}\\
20.53\% &=42.05 \% \times 46.28 \% + 57.94 \% \times 1.86 \%\\
9\% &=16.06 \% \times 49.07 \% + 83.94 \% \times 1.34 \%\\
11.95\% &=20.4 \% \times 50.12 \% + 79.6 \% \times 2.17 \%
\end{split}
\label{EEtest}
\end{equation}

Where $EE_{\rm{a,b,c}}$ the percentage of the light
in the partial simulation ($EE_{\rm{a}} = EE_{\rm{b}}
+ EE_{\rm{c}} = 100 \% $), and $T_{\rm{b,c}}$
the throughput in the partial simulation.

\begin{table}
	\centering
	\caption{Evanescent field contribution results 
    (See section \ref{sec:ev_coupling})}
	\label{table:evanescent}
	\begin{tabular}{cccc}
		\hline
		 & Full field & 
        Cut field & Cut-inside field\\
        \ac{AO} mode &(Throughput)&$(EE_{\rm{b}}
        \times T_{\rm{b}}$)&$(EE_{\rm{c}} \times T_{\rm{c}}$)\\
		\hline \hline
        closed-loop (\%)& 20.53 & 42.05 $\times$ 46.28 & 57.94 $\times$ 1.86 \\
		tip-tilt (\%) & 9 & 16.06 $\times$ 49.07& 83.94 $\times$ 1.34 \\
        open-loop (\%) & 11.95 & 20.4 $\times$ 50.12& 79.6$\times$ 2.17\\
        \hline
	\end{tabular}
\end{table}

The results from this are shown in Table 
\ref{table:evanescent}. This result shows that the
light coupled into the \ac{PD} was not coupled 
entirely at the entrance to the \ac{PD}. We can 
explain this as being due to the small refractive 
index difference between core and cladding 
($\mathrm{\Delta\approx1.76\times10^{-3}}$). This
gives the \ac{PD} a large evanescent field, which
couples light into the waveguides. 

We looked into this further, by examining the 
partial power monitors in RSoft as the light 
propagated along the waveguides. Figure
\ref{partial_ee} shows the normalised power within
the waveguides. As expected, this  drops as 
the light propagates through the \ac{PD}. However
in the second to last section we see the power
increasing slightly. This is due to the power 
monitors in RSoft not taking the evanescent field
of the waveguides into account. As the waveguides
in the second to last section are brought together,
the evanescent field from each one is coupled into
the adjacent waveguide, which means the evanescent
fields overlap, increasing the measured power in 
the \ac{PD}.

To summarise, our findings indicate that up to 2\% of
the light within the slit output originates from evanescent
field coupling.
Thus, a slit mask should be used in front of the \ac{PD}
entrance if the evanescent field is undesired depending
on the scientific goals.

\subsection{Modal noise}

\begin{figure}
\centering
\includegraphics[width=0.5\textwidth]{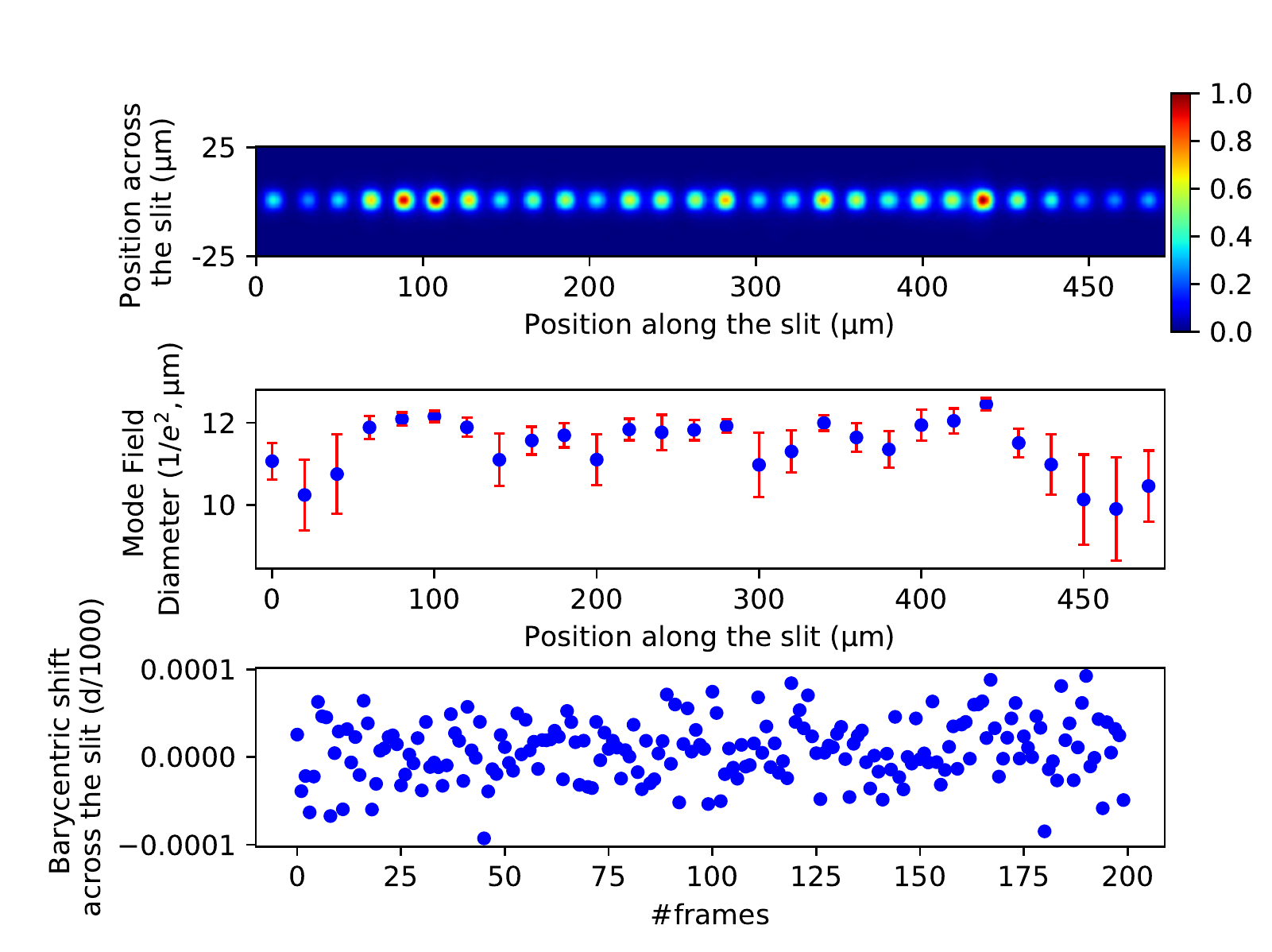}
\caption{Top panel: Near field averaged image 
(intensity) of the separated slit from BeamProp 
simulations. Middle panel: MFD of the slit profile
including 1$\upsigma$ errors from individual frames.
Bottom panel: Measurements of barycentre movement
across the slit from individual frames.}
\label{slit:sep}
\end{figure}

\begin{figure}
\centering
\includegraphics[width=0.5\textwidth]{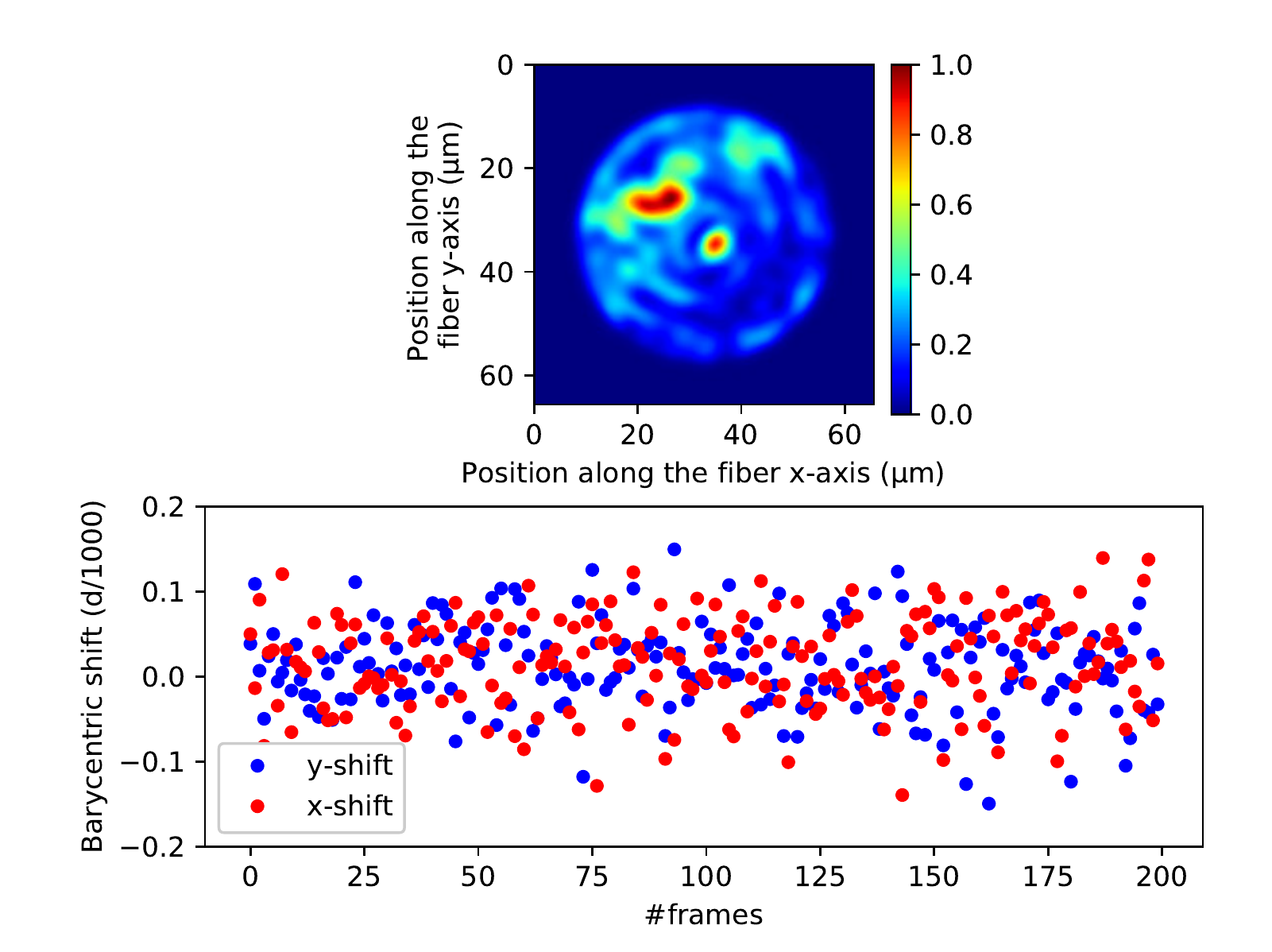}
\caption{Top panel: Typical near field image of the
50 $\upmu$m \ac{MM} fibre output from BeamProp 
simulations. Bottom panel: Measurements of barycentre
movement along the fibre y-axis (blue) and x-axis
(red) from individual frames.}
\label{mm:fiber}
\end{figure}

As we can see from the bottom panel in Figure 
\ref{slit_com} the modal pattern in the slit is not
straight and has some limited residual movement 
even though the slit was configured to be straight.
This, as with modal noise, will limit the spectral 
resolving power of the spectrograph, though not to
the same extent as with the modal noise in conventional
fibres \citep{Chen:2006}. In order to prove that
statement two experiments were performed to justify
our hypothesis. Firstly, following the procedure as
described in section \ref{sec:mnoise} the variation
of the \ac{MFD} and its barycentric position were
calculated for a device identical to the \ac{PD}, 
though at the output level of the slit the waveguides
were separated and not touching each other. Secondly,
the same method was applied to a common circular 
\ac{MM} fibre 50 $\upmu$m in diameter with a 
NA = 0.22 and refractive index of the core equal to
1.45. Results suggest that for the separated version
of the \ac{PD}, barycentric movement is 50\% more 
stable than the original version of the \ac{PD}
(semi amplitude variation $10^{-4}$ ($d$/1000),
see Figure \ref{slit:sep}), while for the \ac{MM}
fibre case the barycentre movement of the average
of the speckles that were calculated, is three
orders larger than the \ac{PD} (semi amplitude
variation $2\times 10^{-2}$ ($d$/1000), see Figure
\ref{mm:fiber}) and qualitatively similar to results
in the literature (e.g. \cite{Feger:2012}).

It should be cautioned that, as noted in 
\cite{Spaleniak:2016} any imperfections in the
manufacture of the slit will result in modal
noise due to movement of the barycentre of the
\ac{MFD}. Following our results above, we would
suggest (as already pointed out in the aforementioned
paper) separated slit cores, to allow reduction of
this modal noise.

We also see no variation in throughput with
wavelength for the \ac{PD}, as seen with similar
devices and wavelength regimes (e.g.
\citep{Spaleniak:2016,Cvetojevic:2017}). This
suggests our device is free of noise caused by
modal mismatch between components (e.g. the
mismatch between a \ac{MM} fibre and \ac{PL} in
\cite{Cvetojevic:2017}). 

\section{Conclusions}
\label{sec:conclusion}

We have conducted a theoretical study concerning
the performance of an existing astrophotonic
component, the \acl{PD}. We make use of Soapy, a
Monte Carlo \ac{AO} simulation program to model
the atmosphere and its impact on the performance
of the device, and BeamProp by RSoft, a 
finite-difference beam propagation solver to 
simulate the device itself. The simulated \ac{AO}
corrected \acp{PSF} were used as an input to our 
replicated \ac{PD} in RSoft.

Our results matched the on-sky results well.
Showing a simulated throughput of 20 $\pm$ 2\% in
closed-loop (compared to the same value on-sky), 9
$\pm$ 2\% in tip-tilt (compared to the same value
on-sky) and 8 $\pm$ 2\% in open-loop (compared to
11 $\pm$ 2\% for on-sky). The slight variation is
likely due to changing atmospheric seeing during
the course of the observations, which were only 
partially reproduced in the simulation. 

We also investigated the effect of modal noise
on the \ac{PD}. We showed that although it is
not completely modal noise free it should show a 
reduction of three orders of magnitude as compared
to a standard \ac{MM} fibre. This can also be
improved by separating the output slit, as
suggested in \cite{Spaleniak:2016}.

Further simulations were used to optimise the 
device and showed a throughput improvement of
6.4\%. This shows the importance of fully
simulating such devices, in particular with
atmospheric effects. 
   
Our simulations also revealed an error in
magnification at the input of the \acl{PD}
reported in \cite{Harris:2015}. A value
of 7.96 arcseconds/mm was reported for the
plate scale while our investigation resulted 
in a plate scale of 6.37 arcseconds/mm.
Optimising this will be important in future 
work for both the devices and also the \acl{AO} 
performance.

Our results suggest that detailed simulations are
a valuable tool for the design of new components
for astronomy with the aim of enabling more precise
measurements, easier calibration of the acquired
data, and more compact instruments for future
telescopes. Simulations like ours can be used to
estimate the on-sky performance in non ideal
observing conditions.

Aims for future work include further optimisation
for better coupling to the telescope \acl{PSF} by
repositioning of the \acl{PD} entrance wave-guide
positions and improvement of the transmission of
the device by a better manufacturing process. 
Additionally, there is high potential for more 
advanced photonic instrument concepts such as an
integrated spatial reformatter feeding an arrayed
waveguide grating (AWG) \citep{Stoll:2017, Cvetojevic:2017}.

\section*{Acknowledgements}

This work was supported by the Deutsche Forschungsgemeinschaft 
(DFG) through project 326946494, `Novel Astronomical
Instrumentation through photonic Reformatting'. Robert J.
Harris is funded/supported by the Carl-Zeiss-Foundation.
R.R.T sincerely thanks the UK Science and Technology
Facilities Council (STFC) for support through an STFC
Consortium Grant (ST/N000625/1).

We would like to thank Dionne M. Haynes from Leibniz
Institute for Astrophysics Potsdam and Ph.D student
Jan Tepper from University of K\"oln for their feedback
improving this study.

This research made use of Astropy, a community-developed
core Python package for Astronomy (Astropy Collaboration, 2013),
Numpy \citep{numpy} and Matplotlib \citep{matplotlib}.


\bibliographystyle{mnras}
\bibliography{references}

\begin{thebibliography}{}
\makeatletter
\relax
\def\mn@urlcharsother{\let\do\@makeother \do\$\do\&\do\#\do\^\do\_\do\%\do\~}
\def\mn@doi{\begingroup\mn@urlcharsother \@ifnextchar [ {\mn@doi@}
  {\mn@doi@[]}}
\def\mn@doi@[#1]#2{\def\@tempa{#1}\ifx\@tempa\@empty \href
  {http://dx.doi.org/#2} {doi:#2}\else \href {http://dx.doi.org/#2} {#1}\fi
  \endgroup}
\def\mn@eprint#1#2{\mn@eprint@#1:#2::\@nil}
\def\mn@eprint@arXiv#1{\href {http://arxiv.org/abs/#1} {{\tt arXiv:#1}}}
\def\mn@eprint@dblp#1{\href {http://dblp.uni-trier.de/rec/bibtex/#1.xml}
  {dblp:#1}}
\def\mn@eprint@#1:#2:#3:#4\@nil{\def\@tempa {#1}\def\@tempb {#2}\def\@tempc
  {#3}\ifx \@tempc \@empty \let \@tempc \@tempb \let \@tempb \@tempa \fi \ifx
  \@tempb \@empty \def\@tempb {arXiv}\fi \@ifundefined
  {mn@eprint@\@tempb}{\@tempb:\@tempc}{\expandafter \expandafter \csname
  mn@eprint@\@tempb\endcsname \expandafter{\@tempc}}}

\bibitem[\protect\citeauthoryear{{Agapito}, {Arcidiacono}, {Quir{\'o}s-Pacheco}
   \& {Esposito}}{{Agapito} et~al.}{2014}]{Agapito:2014}
{Agapito} G.,  {Arcidiacono} C.,  {Quir{\'o}s-Pacheco} F.,   {Esposito} S.,
  2014, \mn@doi [Experimental Astronomy] {10.1007/s10686-014-9380-7}, \href
  {http://adsabs.harvard.edu/abs/2014ExA....37..503A} {37, 503}

\bibitem[\protect\citeauthoryear{{Allington-Smith} et~al.,}{{Allington-Smith}
  et~al.}{2002}]{Allington-Smith:2004}
{Allington-Smith} J.,  et~al., 2002, \mn@doi [\pasp] {10.1086/341712}, \href
  {http://adsabs.harvard.edu/abs/2002PASP..114..892A} {114, 892}

\bibitem[\protect\citeauthoryear{Birks, Gris-S\'{a}nchez, Yerolatsitis,
  Leon-Saval  \& Thomson}{Birks et~al.}{2015}]{Birks:2015}
Birks T.~A.,  Gris-S\'{a}nchez I.,  Yerolatsitis S.,  Leon-Saval S.~G.,
  Thomson R.~R.,  2015, \mn@doi [Adv. Opt. Photon.] {10.1364/AOP.7.000107}, 7,
  107

\bibitem[\protect\citeauthoryear{{Bland-Hawthorn} \& {Horton}}{{Bland-Hawthorn}
  \& {Horton}}{2006}]{Bland-Hawthorn:2006}
{Bland-Hawthorn} J.,  {Horton} A.,  2006, \mn@doi [Proc. SPIE]
  {10.1117/12.670931}, \href
  {http://adsabs.harvard.edu/abs/2006SPIE.6269E..0NB} {6269, 62690N}

\bibitem[\protect\citeauthoryear{{Bland-Hawthorn} et~al.,}{{Bland-Hawthorn}
  et~al.}{2010}]{Bland-Hawthorn:2010}
{Bland-Hawthorn} J.,  et~al., 2010, \mn@doi [Proc. SPIE] {10.1117/12.856347},
  \href {http://adsabs.harvard.edu/abs/2010SPIE.7735E..0NB} {7735, 77350N}

\bibitem[\protect\citeauthoryear{{Bouchy}, {D{\'{\i}}az}, {H{\'e}brard},
  {Arnold}, {Boisse}, {Delfosse}, {Perruchot}  \& {Santerne}}{{Bouchy}
  et~al.}{2013}]{Bouchy:2013}
{Bouchy} F.,  {D{\'{\i}}az} R.~F.,  {H{\'e}brard} G.,  {Arnold} L.,  {Boisse}
  I.,  {Delfosse} X.,  {Perruchot} S.,   {Santerne} A.,  2013, \mn@doi [\aap]
  {10.1051/0004-6361/201219979}, \href
  {http://adsabs.harvard.edu/abs/2013A%26A...549A..49B} {549, A49}

\bibitem[\protect\citeauthoryear{Chen, Reynolds  \& Kost}{Chen
  et~al.}{2006}]{Chen:2006}
Chen C.-H.,  Reynolds R.~O.,   Kost A.,  2006, \mn@doi [Appl. Opt.]
  {10.1364/AO.45.000519}, 45, 519

\bibitem[\protect\citeauthoryear{{Coud{\'e} du Foresto}}{{Coud{\'e} du
  Foresto}}{1994}]{Coude:1994}
{Coud{\'e} du Foresto} V.,  1994, IAU Symposium, \href
  {http://adsabs.harvard.edu/abs/1994IAUS..158..261C} {158, 261}

\bibitem[\protect\citeauthoryear{Crepp}{Crepp}{2014}]{Crepp:2014}
Crepp J.~R.,  2014, \mn@doi [Science] {10.1126/science.1262071}, 346, 809

\bibitem[\protect\citeauthoryear{{Cunningham}}{{Cunningham}}{2009}]{Cunningham:2009}
{Cunningham} C.,  2009, \mn@doi [Nature Photonics] {10.1038/nphoton.2009.49},
  \href {http://adsabs.harvard.edu/abs/2009NaPho...3..239C} {3, 239}

\bibitem[\protect\citeauthoryear{Cvetojevic, Lawrence, Ellis, Bland-Hawthorn,
  Haynes  \& Horton}{Cvetojevic et~al.}{2009}]{Cvetojevic:2009}
Cvetojevic N.,  Lawrence J.~S.,  Ellis S.~C.,  Bland-Hawthorn J.,  Haynes R.,
  Horton A.,  2009, \mn@doi [Opt. Express] {10.1364/OE.17.018643}, 17, 18643

\bibitem[\protect\citeauthoryear{Cvetojevic, Jovanovic, Lawrence, Withford  \&
  Bland-Hawthorn}{Cvetojevic et~al.}{2012}]{Cvetojevic:2012}
Cvetojevic N.,  Jovanovic N.,  Lawrence J.,  Withford M.,   Bland-Hawthorn J.,
  2012, \mn@doi [Opt. Express] {10.1364/OE.20.002062}, 20, 2062

\bibitem[\protect\citeauthoryear{{Cvetojevic} et~al.,}{{Cvetojevic}
  et~al.}{2017}]{Cvetojevic:2017}
{Cvetojevic} N.,  et~al., 2017, \mn@doi [Opt. Express] {10.1364/OE.25.025546},
  25, 25546

\bibitem[\protect\citeauthoryear{{Dekany} et~al.,}{{Dekany}
  et~al.}{2013}]{Dekany:2013}
{Dekany} R.,  et~al., 2013, \mn@doi [\apj] {10.1088/0004-637X/776/2/130}, \href
  {http://adsabs.harvard.edu/abs/2013ApJ...776..130D} {776, 130}

\bibitem[\protect\citeauthoryear{{Feger}, {Brucalassi}, {Grupp}, {Lang-Bardl},
  {Holzwarth}, {Hopp}  \& {Bender}}{{Feger} et~al.}{2012}]{Feger:2012}
{Feger} T.,  {Brucalassi} A.,  {Grupp} F.~U.,  {Lang-Bardl} F.,  {Holzwarth}
  R.,  {Hopp} U.,   {Bender} R.,  2012, \mn@doi [Proc. SPIE]
  {10.1117/12.925624}, \href
  {http://adsabs.harvard.edu/abs/2012SPIE.8446E..92F} {8446, 844692}

\bibitem[\protect\citeauthoryear{Halverson, Roy, Mahadevan  \&
  Schwab}{Halverson et~al.}{2015}]{Halverson:2015}
Halverson S.,  Roy A.,  Mahadevan S.,   Schwab C.,  2015, \mn@doi [\apjl]
  {10.1088/2041-8205/814/2/L22}, 814, L22

\bibitem[\protect\citeauthoryear{{Harris} \& {Allington-Smith}}{{Harris} \&
  {Allington-Smith}}{2013}]{Harris:2013}
{Harris} R.~J.,  {Allington-Smith} J.~R.,  2013, \mn@doi [\mnras]
  {10.1093/mnras/sts265}, \href
  {http://adsabs.harvard.edu/abs/2013MNRAS.428.3139H} {428, 3139}

\bibitem[\protect\citeauthoryear{Harris et~al.,}{Harris
  et~al.}{2015}]{Harris:2015}
Harris R.~J.,  et~al., 2015, \mn@doi [\mnras] {10.1093/mnras/stv410}, 450, 428

\bibitem[\protect\citeauthoryear{{Hook}, {J{\o}rgensen}, {Allington-Smith},
  {Davies}, {Metcalfe}, {Murowinski}  \& {Crampton}}{{Hook}
  et~al.}{2004}]{Hook:2004}
{Hook} I.~M.,  {J{\o}rgensen} I.,  {Allington-Smith} J.~R.,  {Davies} R.~L.,
  {Metcalfe} N.,  {Murowinski} R.~G.,   {Crampton} D.,  2004, \mn@doi [\pasp]
  {10.1086/383624}, \href {http://adsabs.harvard.edu/abs/2004PASP..116..425H}
  {116, 425}

\bibitem[\protect\citeauthoryear{Hunter}{Hunter}{2007}]{matplotlib}
Hunter J.~D.,  2007, \mn@doi [Computing In Science \& Engineering]
  {10.1109/MCSE.2007.55}, 9, 90

\bibitem[\protect\citeauthoryear{{Iuzzolino}, {Tozzi}, {Sanna}, {Zangrilli}  \&
  {Oliva}}{{Iuzzolino} et~al.}{2014}]{Iuzzolino:2014}
{Iuzzolino} M.,  {Tozzi} A.,  {Sanna} N.,  {Zangrilli} L.,   {Oliva} E.,  2014,
  \mn@doi [Proc. SPIE] {10.1117/12.2055093}, \href
  {http://adsabs.harvard.edu/abs/2014SPIE.9147E..66I} {9147, 914766}

\bibitem[\protect\citeauthoryear{{Jovanovic} et~al.,}{{Jovanovic}
  et~al.}{2015}]{Jovanovic:2015}
{Jovanovic} N.,  et~al., 2015, \mn@doi [\pasp] {10.1086/682989}, \href
  {http://adsabs.harvard.edu/abs/2015PASP..127..890J} {127, 890}

\bibitem[\protect\citeauthoryear{{Jovanovic}, {Schwab}, {Cvetojevic}, {Guyon}
  \& {Martinache}}{{Jovanovic} et~al.}{2016}]{Jovanovic:2016}
{Jovanovic} N.,  {Schwab} C.,  {Cvetojevic} N.,  {Guyon} O.,   {Martinache} F.,
   2016, \mn@doi [\pasp] {10.1088/1538-3873/128/970/121001}, \href
  {http://adsabs.harvard.edu/abs/2016PASP..128l1001J} {128, 121001}

\bibitem[\protect\citeauthoryear{Lemke, Corbett, Allington-Smith  \&
  Murray}{Lemke et~al.}{2011}]{Lemke:2011}
Lemke U.,  Corbett J.,  Allington-Smith J.,   Murray G.,  2011, \mn@doi
  [\mnras] {10.1111/j.1365-2966.2011.19312.x}, 417, 689

\bibitem[\protect\citeauthoryear{Leon-Saval, Birks, Bland-Hawthorn  \&
  Englund}{Leon-Saval et~al.}{2005}]{Leon-Saval:2005}
Leon-Saval S.~G.,  Birks T.~A.,  Bland-Hawthorn J.,   Englund M.,  2005, in
  Optical Fiber Communication Conference and Exposition and The National Fiber
  Optic Engineers Conference. Optical Society of America, p. PDP25, \url
  {http://www.osapublishing.org/abstract.cfm?URI=OFC-2005-PDP25}

\bibitem[\protect\citeauthoryear{{Leon-Saval}, {Betters}  \&
  {Bland-Hawthorn}}{{Leon-Saval} et~al.}{2012}]{Leon-Saval:2012}
{Leon-Saval} S.~G.,  {Betters} C.~H.,   {Bland-Hawthorn} J.,  2012, \mn@doi
  [Proc. SPIE] {10.1117/12.925254}, \href
  {http://adsabs.harvard.edu/abs/2012SPIE.8450E..1KL} {8450, 84501K}

\bibitem[\protect\citeauthoryear{{Leon-Saval}, {Argyros}  \&
  {Bland-Hawthorn}}{{Leon-Saval} et~al.}{2013}]{Leon-Saval:2013}
{Leon-Saval} S.~G.,  {Argyros} A.,   {Bland-Hawthorn} J.,  2013, \mn@doi
  [Nanophotonics] {10.1515/nanoph-2013-0035}, \href
  {http://adsabs.harvard.edu/abs/2013Nanop...2..429L} {2, 429}

\bibitem[\protect\citeauthoryear{{MacLachlan}, {Harris}, {Choudhury},
  {Arriola}, {Brown}, {Allington-Smith}  \& {Thomson}}{{MacLachlan}
  et~al.}{2014}]{MacLachlan:2014}
{MacLachlan} D.~G.,  {Harris} R.,  {Choudhury} D.,  {Arriola} A.,  {Brown} G.,
  {Allington-Smith} J.,   {Thomson} R.~R.,  2014, \mn@doi [Proc. SPIE]
  {10.1117/12.2055904}, \href
  {http://adsabs.harvard.edu/abs/2014SPIE.9151E..1WM} {9151, 91511W}

\bibitem[\protect\citeauthoryear{{MacLachlan} et~al.,}{{MacLachlan}
  et~al.}{2017}]{MacLachlan:2017}
{MacLachlan} D.~G.,  et~al., 2017, \mn@doi [\mnras] {10.1093/mnras/stw2558},
  \href {http://adsabs.harvard.edu/abs/2017MNRAS.464.4950M} {464, 4950}

\bibitem[\protect\citeauthoryear{{Macintosh} et~al.,}{{Macintosh}
  et~al.}{2014}]{Macintosh:2014}
{Macintosh} B.,  et~al., 2014, \mn@doi [Proceedings of the National Academy of
  Science] {10.1073/pnas.1304215111}, \href
  {http://adsabs.harvard.edu/abs/2014PNAS..11112661M} {111, 12661}

\bibitem[\protect\citeauthoryear{{Mayor} et~al.,}{{Mayor}
  et~al.}{2003}]{Mayor:2003}
{Mayor} M.,  et~al., 2003, The Messenger, \href
  {http://adsabs.harvard.edu/abs/2003Msngr.114...20M} {114, 20}

\bibitem[\protect\citeauthoryear{{McCoy}, {Ramsey}, {Mahadevan}, {Halverson}
  \& {Redman}}{{McCoy} et~al.}{2012}]{McCoy:2012}
{McCoy} K.~S.,  {Ramsey} L.,  {Mahadevan} S.,  {Halverson} S.,   {Redman}
  S.~L.,  2012, \mn@doi [Proc. SPIE] {10.1117/12.926287}, \href
  {http://adsabs.harvard.edu/abs/2012SPIE.8446E..8JM} {8446, 84468J}

\bibitem[\protect\citeauthoryear{{Mueller} et~al.,}{{Mueller}
  et~al.}{2014}]{Mueller:2014}
{Mueller} M.,  et~al., 2014, \mn@doi [Proc. SPIE] {10.1117/12.2056440}, \href
  {http://adsabs.harvard.edu/abs/2014SPIE.9147E..9AM} {9147, 91479A}

\bibitem[\protect\citeauthoryear{Myers et~al.,}{Myers
  et~al.}{2008}]{Myers:2008}
Myers R.~M.,  et~al., 2008, \mn@doi [Proc. SPIE] {10.1117/12.789544}, 7015,
  70150E

\bibitem[\protect\citeauthoryear{Nasu, Kohtoku  \& Hibino}{Nasu
  et~al.}{2005}]{Nasu:2005}
Nasu Y.,  Kohtoku M.,   Hibino Y.,  2005, \mn@doi [Opt. Lett.]
  {10.1364/OL.30.000723}, 30, 723

\bibitem[\protect\citeauthoryear{Noguchi et~al.,}{Noguchi
  et~al.}{2002}]{Noguchi:2002}
Noguchi K.,  et~al., 2002, \mn@doi [Publications of the Astronomical Society of
  Japan] {10.1093/pasj/54.6.855}, 54, 855

\bibitem[\protect\citeauthoryear{{Perruchot} et~al.,}{{Perruchot}
  et~al.}{2011}]{Perruchot:2011}
{Perruchot} S.,  et~al., 2011, \mn@doi [Proc. SPIE] {10.1117/12.892466}, \href
  {http://adsabs.harvard.edu/abs/2011SPIE.8151E..15P} {8151, 815115}

\bibitem[\protect\citeauthoryear{{Probst} et~al.,}{{Probst}
  et~al.}{2015}]{Probst:2015}
{Probst} R.~A.,  et~al., 2015, \mn@doi [New Journal of Physics]
  {10.1088/1367-2630/17/2/023048}, \href
  {http://adsabs.harvard.edu/abs/2015NJPh...17b3048P} {17, 023048}

\bibitem[\protect\citeauthoryear{{Quirrenbach} et~al.,}{{Quirrenbach}
  et~al.}{2016}]{Quirrenbach:2016}
{Quirrenbach} A.,  et~al., 2016, \mn@doi [Proc. SPIE] {10.1117/12.2231880},
  \href {http://adsabs.harvard.edu/abs/2016SPIE.9908E..12Q} {9908, 990812}

\bibitem[\protect\citeauthoryear{Rawson, Goodman  \& Norton}{Rawson
  et~al.}{1980}]{Rawson:1980}
Rawson E.~G.,  Goodman J.~W.,   Norton R.~E.,  1980, \mn@doi [J. Opt. Soc. Am.]
  {10.1364/JOSA.70.000968}, 70, 968

\bibitem[\protect\citeauthoryear{{Reeves}}{{Reeves}}{2016}]{Reeves:2016}
{Reeves} A.,  2016, \mn@doi [Proc. SPIE] {10.1117/12.2232438}, \href
  {http://adsabs.harvard.edu/abs/2016SPIE.9909E..7FR} {9909, 99097F}

\bibitem[\protect\citeauthoryear{{Schwab}, {Leon-Saval}, {Betters},
  {Bland-Hawthorn}  \& {Mahadevan}}{{Schwab} et~al.}{2014}]{Schwab:2014}
{Schwab} C.,  {Leon-Saval} S.~G.,  {Betters} C.~H.,  {Bland-Hawthorn} J.,
  {Mahadevan} S.,  2014, \mn@doi [IAU Symposium] {10.1017/S1743921313013264},
  \href {http://adsabs.harvard.edu/abs/2014IAUS..293..403S} {293, 403}

\bibitem[\protect\citeauthoryear{{Spaleniak}, {Jovanovic}, {Gross}, {Ireland},
  {Lawrence}  \& {Withford}}{{Spaleniak} et~al.}{2013}]{Spaleniak:2013}
{Spaleniak} I.,  {Jovanovic} N.,  {Gross} S.,  {Ireland} M.~J.,  {Lawrence}
  J.~S.,   {Withford} M.~J.,  2013, \mn@doi [Optics Express]
  {10.1364/OE.21.027197}, \href
  {http://adsabs.harvard.edu/abs/2013OExpr..2127197S} {21, 27197}

\bibitem[\protect\citeauthoryear{{Spaleniak} et~al.,}{{Spaleniak}
  et~al.}{2016}]{Spaleniak:2016}
{Spaleniak} I.,  et~al., 2016, \mn@doi [Proc. SPIE] {10.1117/12.2232708}, \href
  {http://adsabs.harvard.edu/abs/2016SPIE.9912E..28S} {9912, 991228}

\bibitem[\protect\citeauthoryear{Stoll, Zhang, Haynes  \& Roth}{Stoll
  et~al.}{2017}]{Stoll:2017}
Stoll A.,  Zhang Z.,  Haynes R.,   Roth M.,  2017, \mn@doi [Photonics]
  {10.3390/photonics4020030}, 4

\bibitem[\protect\citeauthoryear{Synopsys}{Synopsys}{}]{rsoft}
Synopsys, RSoft Photonic System Design Suite Version 2017.03, \url
  {https://optics.synopsys.com/rsoft/}

\bibitem[\protect\citeauthoryear{{Thomson}, {Birks}, {Leon-Saval}, {Kar}  \&
  {Bland-Hawthorn}}{{Thomson} et~al.}{2011}]{Thomson:2011}
{Thomson} R.,  {Birks} T.,  {Leon-Saval} S.,  {Kar} A.,   {Bland-Hawthorn} J.,
  2011, \mn@doi [Optics Express] {10.1364/OE.19.005698}, \href
  {http://adsabs.harvard.edu/abs/2011OExpr..19.5698T} {19, 5698}

\bibitem[\protect\citeauthoryear{{Tollestrup}, {Pazder}, {Barrick}, {Martioli},
  {Schiavon}, {Anthony}, {Halman}  \& {Veillet}}{{Tollestrup}
  et~al.}{2012}]{Tollestrup:2012}
{Tollestrup} E.~V.,  {Pazder} J.,  {Barrick} G.,  {Martioli} E.,  {Schiavon}
  R.,  {Anthony} A.,  {Halman} M.,   {Veillet} C.,  2012, \mn@doi [Proc. SPIE]
  {10.1117/12.926626}, \href
  {http://adsabs.harvard.edu/abs/2012SPIE.8446E..2AT} {8446, 84462A}

\bibitem[\protect\citeauthoryear{Van~der walt, Colbert  \& Ga{\"{e}}l}{Van~der
  walt et~al.}{2011}]{numpy}
Van~der walt S.,  Colbert S.~C.,   Ga{\"{e}}l V.,  2011, Computing in Science
  {\&} Engineering, 13, 22

\bibitem[\protect\citeauthoryear{{Vogt} et~al.,}{{Vogt}
  et~al.}{1994}]{Vogt:1994}
{Vogt} S.~S.,  et~al., 1994, \mn@doi [Proc. SPIE] {10.1117/12.176725}, \href
  {http://adsabs.harvard.edu/abs/1994SPIE.2198..362V} {2198, 362}

\bibitem[\protect\citeauthoryear{{Weitzel}, {Krabbe}, {Kroker}, {Thatte},
  {Tacconi-Garman}, {Cameron}  \& {Genzel}}{{Weitzel}
  et~al.}{1996}]{Weitzel:1996}
{Weitzel} L.,  {Krabbe} A.,  {Kroker} H.,  {Thatte} N.,  {Tacconi-Garman}
  L.~E.,  {Cameron} M.,   {Genzel} R.,  1996, \aaps, \href
  {http://adsabs.harvard.edu/abs/1996A%26AS..119..531W} {119, 531}

\bibitem[\protect\citeauthoryear{{Yerolatsitis}, {Harrington}  \&
  {Birks}}{{Yerolatsitis} et~al.}{2017}]{Yerolatsitis:2017}
{Yerolatsitis} S.,  {Harrington} K.,   {Birks} T.~A.,  2017, \mn@doi [Optics
  Express] {10.1364/OE.25.018713}, \href
  {http://adsabs.harvard.edu/abs/2017OExpr..2518713Y} {25, 18713}

\bibitem[\protect\citeauthoryear{{Zerbi} et~al.,}{{Zerbi}
  et~al.}{2014}]{Zerbi:2014}
{Zerbi} F.~M.,  et~al., 2014, \mn@doi [Proc. SPIE] {10.1117/12.2055329}, \href
  {http://adsabs.harvard.edu/abs/2014SPIE.9147E..23Z} {9147, 914723}

\makeatother
\end{thebibliography}


\appendix




 \bsp	
\label{lastpage}
\end{document}